\documentclass[12pt,draftcls,a4paper,onecolumn,journal]{IEEEtran}

\usepackage{amsmath,graphicx,subfigure,epstopdf,multirow,color,flushend,cite,url,slashbox,placeins,footnote}
\usepackage[normalem]{ulem}
\newlength\savedwidth
\newcommand{\whline}{\noalign{\global\savedwidth\arrayrulewidth
    \global\arrayrulewidth 1.2pt} \hline
  \noalign{\global\arrayrulewidth\savedwidth} }

\makeatletter
\def\ps@headings{%
\def\@oddhead{\mbox{}\footnotesize\rightmark \hfil\thepage}%
\def\@evenhead{\footnotesize\thepage \hfil \thepage\leftmark\mbox{}}%
}
\makeatother

\renewcommand{\leftmark}{\underline{This article has been submitted for possible publication in IEEE Transactions on Wireless Communications}}
\renewcommand{\rightmark}{\underline{This article has been submitted for possible publication in IEEE Transactions on Wireless Communications}}

\begin{document}

\title{Coexistence and Interference Mitigation for Wireless Body Area Networks: Improvements using On-Body Opportunistic Relaying}

\author{Jie~Dong,~\IEEEmembership{Student~Member,~IEEE,}
        David~Smith,~\IEEEmembership{Member,~IEEE,\\}
\IEEEauthorblockA{National ICT Australia (NICTA)$^\dag$ and The Australian National University\\
(e-mail: Jie.Dong@nicta.com.au)}}

\maketitle

\begin{abstract}
   Coexistence, and hence interference mitigation, across multiple wireless body area networks (WBANs) is an important problem as WBANs become more pervasive. Here, two-hop relay-assisted cooperative communications using opportunistic relaying (OR) is proposed for enhancement of coexistence for WBANs. Suitable time division multiple access (TDMA) schemes are employed for both intra-WBAN and inter-WBANs access protocols. To emulate actual conditions of WBAN use, extensive on-body and inter-body ``everyday" channel measurements are employed. In addition, a realistic inter-WBAN channel model is simulated to investigate the effect of body shadowing and hub location on WBAN performance in enabling coexistence. When compared with single-link communications, it is found that opportunistic relaying can provide significant improvement, in terms of signal-to-interference+noise ratio (SINR), to outage probability and level crossing rate (LCR) into outages. However, average outage duration (AOD) is not affected. In addition a lognormal distribution shows a better fit to received SINR when the channel coherence time is small, and a Nakagami-m distribution is a more common fit when the channel is more stable. According to the estimated SINR distributions, theoretical outage probability, LCR and AOD are shown to match the empirical results well.
\end{abstract}


\newpage
\section{Introduction}

\IEEEPARstart{T}{he} rapid development of semiconductor technology and wireless communications has led to a new generation of wireless sensor networks, wireless body area networks (WBANs), which allow inexpensive and continuous monitoring of human physiological condition and activity, with real-time updates \cite{Astrin2007}. Sensors can be implanted inside or placed on the human body without causing inconvenience and impairing daily activities. Compared with traditional monitoring, WBAN can provide flexibility and location independency \cite{Lewis2008}. In 2009, there were approximately 11 million active WBAN units around the world, and this number was predicted to reach 420 million by 2014 \cite{ABIResearch}.


A traditional WBAN has a single-link star topology, consisting of several sensors and a hub (i.e., gateway device) \cite{tg6_d}. According to the WBAN’s application requirements, it is essential to maintain high communications reliability at the same time as minimizing power consumption. Therefore, in order to overcome large path losses typically experienced in single-link, star topology, WBAN communications, two-hop cooperative communications is an alternative implementation in the IEEE 802.15.6 WBAN standard \cite{tg6_d}. In both narrow-band \cite{Dong2011,SMITH:WCNC:2012,Ferrand:AT:2011,Dong2012,Dong:ICC:2013} and ultra-wideband WBAN systems \cite{Chen09JSAC}, some two-hop cooperative communication schemes have been investigated that provide significant performance improvement. In \cite{Dong2012} idle sensors are used as relays and selection combining (SC) is used to provide diversity gain at the hub. When using either maximal ratio combining (MRC) or SC with decode-and-forward relay communications, it shows an up-to-14 dB increase in  bit error probability, when employing an IEEE 802.15.6 compliant BCH coding and GFSK modulation scheme \cite{Dong2011}.

Importantly, the improvement in \cite{Dong2011} was determined for an isolated WBAN. However, in a lot of circumstances, the WBAN will need to perform reliably in the vicinity of other WBANs such as in the foyer of a medical center. In addition, as a WBAN may be highly mobile, it is generally not feasible to globally coordinate multiple WBANs\cite{Kim2012}. Therefore, inter-WBAN interference may lead to unreliable communications. Furthermore, according to specifications for the IEEE 802.15.6 standard, the performance of a WBAN should remain reliable when up-to 10 WBANs are co-located in a 6-by-6-by-6m space \cite{TRD}. In light of all these considerations for WBANs this paper focuses on interference mitigation and hence coexistence enhancement by using simple two-hop, relay-assisted cooperative communications.

The contributions of this paper are as follows:
\begin{enumerate}
\item
Multiple WBANs are simulated employing extensive ``everyday" on-body and inter-body channel gain measurements\cite{NICTAdata}, with the WBAN-of-interest consisting of one hub, two possible relays/sensors and eight other sensors. Within each WBAN, the central hub coordinates associated sensors using time division multiple access (TDMA). And, in accordance with an option for physical layer transmission in the IEEE 802.15.6 BAN standard~\cite{tg6_d}, sensors use BCH coded GFSK modulation to transmit packets at 2.4 GHz carrier frequency.
\item
In this context, two non-coordinated neighboring WBANs are modeled to investigate the WBANs coexistence. Modified TDMA is employed as an inter-WBAN access scheme, due to good performance in interference mitigation with respect to channel quality and low power consumption \cite{Zhang2010};
\item
Two-hop cooperative communications is employed for the WBAN-of-interest. Two different implementations are investigated. In the first implementation, the choice of relay node changes according to an activation of any of the sensors as relays\cite{Dong2012}. Whereas, two fixed relays attached to the left hip and right hip respectively are introduced in the second implementation \cite{Dong:ICC:2013}. In this paper, we will primarily focus on the second implementation.
\item
Three-branch opportunistic relaying (OR), similar to using selection combining, is used as it has low complexity and low power consumption. In OR, only a single relay with the most reliable signal path to the destination forwards a packet per hop. This OR is subsequently shown to have significant performance advantages over single-hop communications for both first and second-order outage statistics in the presence of interference, both theoretically and empirically in terms of signal-to-interference-plus-noise-ratio (SINR)
\end{enumerate}

The performance of closely-located WBANs will be interference limited. Therefore, the effectiveness of two-hop communications scheme using relays on interference mitigation is studied based on signal-to-interference-plus-noise-ratio (SINR) of each received packet. Both the first order statistics of outage probability and second order statistics of level crossing rate (LCR) and average outage duration (AOD) for SINR are derived in this paper. The results are compared with respect to the coexistence of two traditional star-topology, single-hop WBANs with the same channel data and the same TDMA schemes used. Here the work published in \cite{Dong2012} and \cite{Dong:ICC:2013} is combined and extended, the extension including analysis of AOD. In other important extensions of \cite{Dong2012},\cite{Dong:ICC:2013}, the best, simple distribution of SINR values is found according to maximum-likelihood parameter estimation; and based on the SINR distributions derived, we generate theoretical outage probability, level crossing rates and average outage duration and compare them with experimental results.

The rest of the paper is organized as follows. In Section II, details of the system model used are given, including two different implementations of cooperative communications. Then Section III provides the analysis of experimental results for WBANs where relay positions are varied over the entire body. In Section IV, analysis of cooperative communications in the WBAN-of-interest using two set relay positions are presented. First order statistics, outage probability and second order statistics, level crossing rate and average outage duration, are derived from contiguous empirical SINR values and are compared with theory according to fitted distributions . In Section V, coexistence performance of WBANs employing two-hop opportunistic relaying cooperative communications with fixed relays is compared with star topology single-link communications. It's effectiveness is shown based on a comparison of outage probability, level crossing rate and average outage duration of the two schemes. Section VI provides some concluding remarks on coexistence between non-coordinated WBANs being better facilitated by cooperative communications on any given WBAN.

\section{System Model}
As mentioned in the introduction, there are two different dual-hop relay-assisted cooperative communications schemes implemented. In this section, we first present common configurations shared by both implementations, which includes intra- and inter-WBAN access schemes, coding and modulation, and opportunistic relaying. Then their differences in relay selection, and intra- and inter-WBAN channel model, together with the techniques for overlaying the models, are explained.

\subsection{Common Setup}
\subsubsection{Intra-WBAN Access Scheme}
\begin{figure}
\centering
\includegraphics[width=0.8\columnwidth]{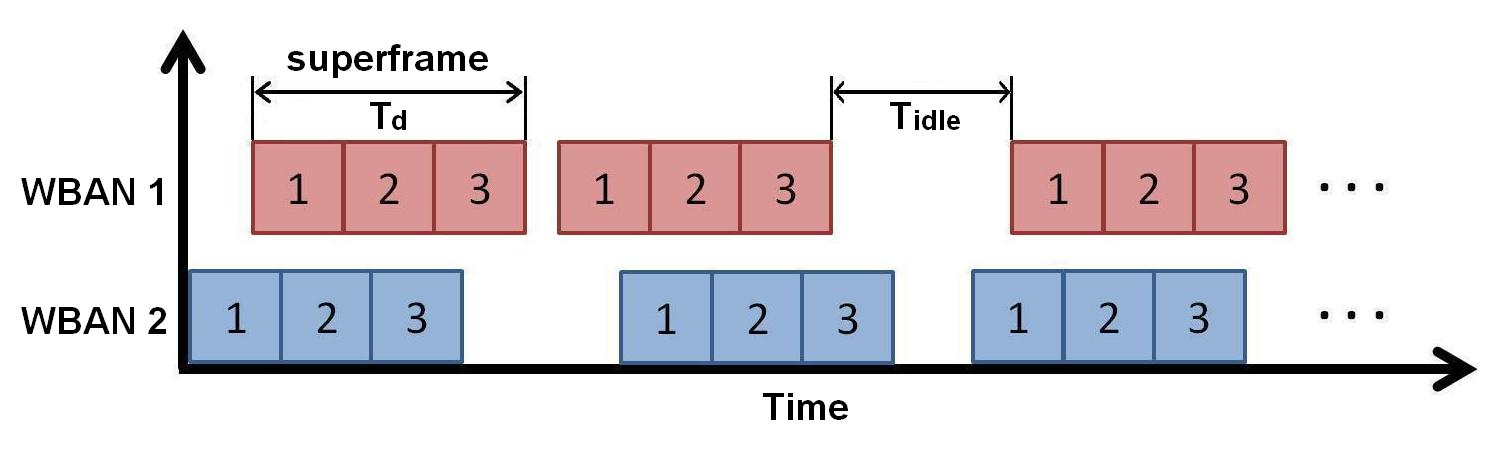}
\caption{Intra-WBAN and Inter-WBANs TDMA scheme}
\label{fig: TDMA}
\end{figure}

The basic model of a WBAN used in the simulation consists of one hub (gateway) and three active sensor nodes. In this star topology system, the hub coordinates the sensors with a time division multiple access (TDMA) scheme. Therefore, as soon as the hub broadcasts a beacon signal, nodes respond by transmitting the collected information back according to a pre-defined sequence for their allocated time slots. This process is shown in Fig. \ref{fig: TDMA}, in which the labels indicate node numbers. In Fig. \ref{fig: TDMA}, the beacon signal is neglected since it is so short compared to the time-length of transmissions by nodes. In terms of two-hop cooperative communications, two relays are chosen from at the hub. On-body hub and relay locations and the way to select active relays will be explained in Section \ref{sec:relset}.

Here we denote the time period starting from beacon signal to all nodes finishing transmission as a superframe, and this concept will be used in latter sections. In analysis and simulation, it is assumed that each sensor only transmits one packet of information in a single superframe. After receiving information packets from all sensors, the WBAN goes into idle mode and waits until the next beacon period starts. The length of the idle mode depends on the inter-WBAN access scheme.

\subsubsection{Inter-WBAN Access Scheme}
In consideration of co-channel interference mitigation and power consumption reduction, TDMA is employed as a co-channel access scheme across all WBANs. Assume there are $N_c$ WBANs located in close proximity and the number is fixed during the period of simulation, then the channel is evenly divided into $N_c$ time slots. Each time slot has a superframe length $T_d$, which allows every WBAN to collect all information packets from sensors in its own network. Therefore, under this TDMA scheme, a WBAN is required to wait for time $(N_c-1) \times T_d$ for other co-located systems to complete their operations before its next transmission cycle starts. This waiting period is also called an idle period, which is denoted as $T_{idle}$. However, because global inter-network coordination is generally not feasible, the execution of this scheme is different from a traditional implementation of TDMA. As shown in Fig.~\ref{fig: TDMA}, a WBAN chooses the start time of every superframe randomly, following a uniform distribution over $[0, T_d+T_{idle}]$. In this paper, inter-WBAN TDMA is used across two WBANs employing this configuration, but this can be generalized to any number of closely located WBANs.

\subsubsection{Coding and Modulation Scheme}
In the IEEE 802.15.6 BAN standard \cite{tg6_d}, BCH coding together with GFSK modulation at 2.4 GHz carrier frequency is specified as an option for physical layer transmission. For this option sensors encode their messages with BCH(31,19), which is a shortened version of BCH(63,51). These encoded messages are then modulated prior to transmission using Gaussian-minimum-shift-keying (GMSK), which is a special type of Gaussian-FSK (GFSK) with a modulation index of 0.5.

\subsubsection{Opportunistic Relaying}
As mentioned in the introduction, two relays are potentially chosen from while each sensor transmits. Therefore, there are three paths that the signal can take from sensor to hub. One is the direct link from the active sensor to the hub, while the other two are decode-and-forward links via either relay. Here, a three-branch opportunistic relaying (OR) scheme is implemented similar to using selection combining (SC) at the hub. The difference between OR and SC is that in OR, the best path is chosen prior to the sensor transmission and only the best path is activated. However, in SC, signals are sent over all three links, and the hub receives three copies of the signal, and the one with the best quality is subsequently chosen. In this way, OR reduces power consumption and mitigates interference with other co-located WBANs. In this paper, the channel quality of different paths is estimated using SINR of packets received at relay and hub nodes. SINR $\nu$ at every node is defined as:

\begin{equation}
\nu = \frac{a_{sig}|h_{TxRx}|^2}{|\varepsilon_{noise}|^2 + \sum(a_{int,i}|h_{int,i}|^2)}~,
\label{equ:SINR}
\end{equation}
where $a_{sig}$ and $a_{int}$ represent the transmit power of a packet from WBAN-of-interest and interfering WBAN respectively; $|h_{TxRx}|$ and $|h_{int}|$ represent the average channel gain for the source to destination and interfering source to destination links across the time duration of the transmitted signal packet. The subscript $i$ indicates the $i$th interfering source. $|\varepsilon_{noise}|$ is the average noise power experienced at the node-of-interest across the time duration of the transmitted signal packet. Throughout this paper, normally distributed additive white Gaussian noise is assumed, with $|\varepsilon_{noise}| = -95$~dBm in all analysis\footnote{Where we note that $-95$~dBm corresponds to a receiver sensitivity specified for 2.4~GHz carrier frequency in IEEE 802.15.6 \cite{tg6_d}.}.

In terms of two relay links, the channel quality is evaluated according to the minimal SINR among two hops at the beginning of a superframe. The selected path $\mathrm{j}_{OR}$ is then determined by the following algorithm:

\begin{align}
\mathrm{j}_{OR} &= \arg \max[ \nu_{1}, \nu_{2}, \nu_{sh} ],\nonumber \\
\nu_{k} &= \min[\nu_{r_k h}, \nu_{sr_k}],\:\textrm{with}\: k = 1\:\textrm{or}\:2.
\label{equ:OR}
\end{align}

The subscripts $s,~r_k~and~h$ indicate sensor, $k$th relay and the hub respectively; $\nu_{ab}$ represents the SINR of received packet transmitted from node $a$ to node $b$.

\subsection{Types of Relay Selection and WBAN Channel Models}
\label{sec:relset}
\subsubsection{Relay Selection}
\begin{figure}[]
\centering
\subfigure[Varying Relays Implementation]
{\includegraphics[width=0.8\columnwidth]{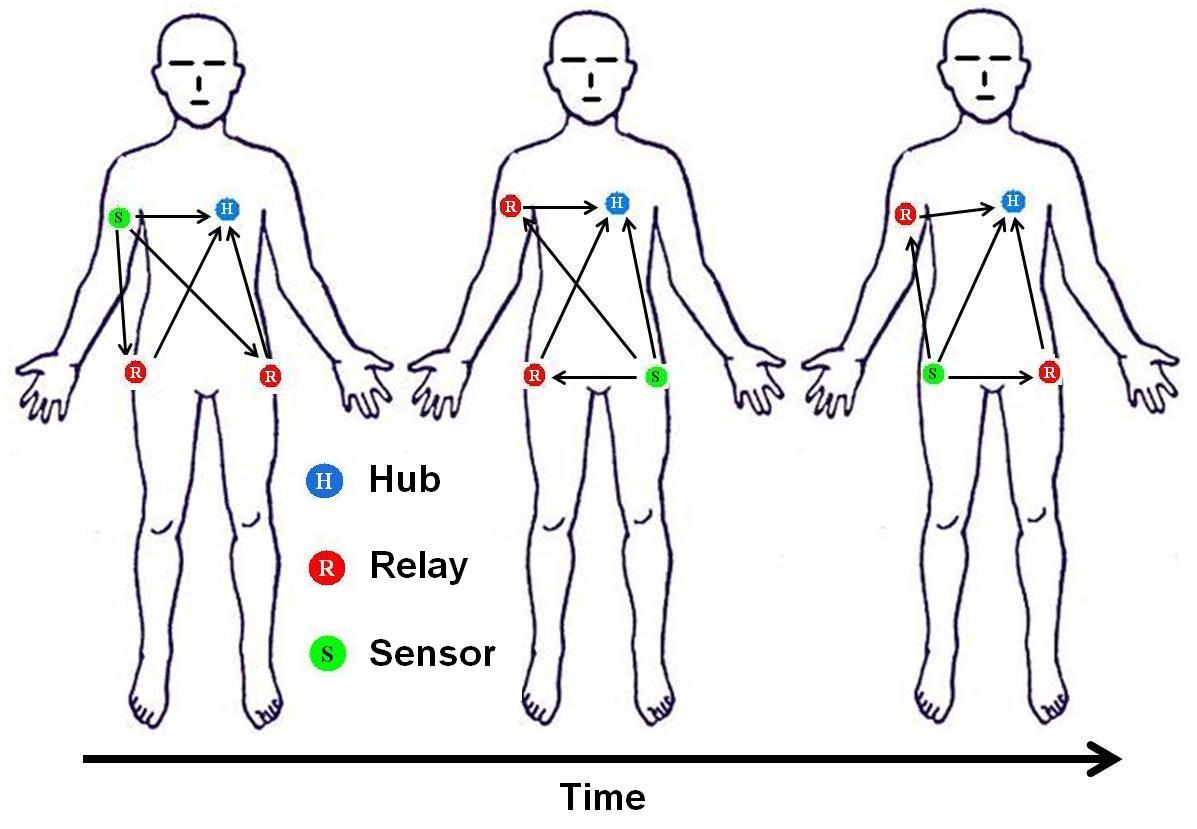}
\label{fig:Varying Relays Implementation}}
\subfigure[Fixed Relays Implementation]
{\includegraphics[width=0.8\columnwidth]{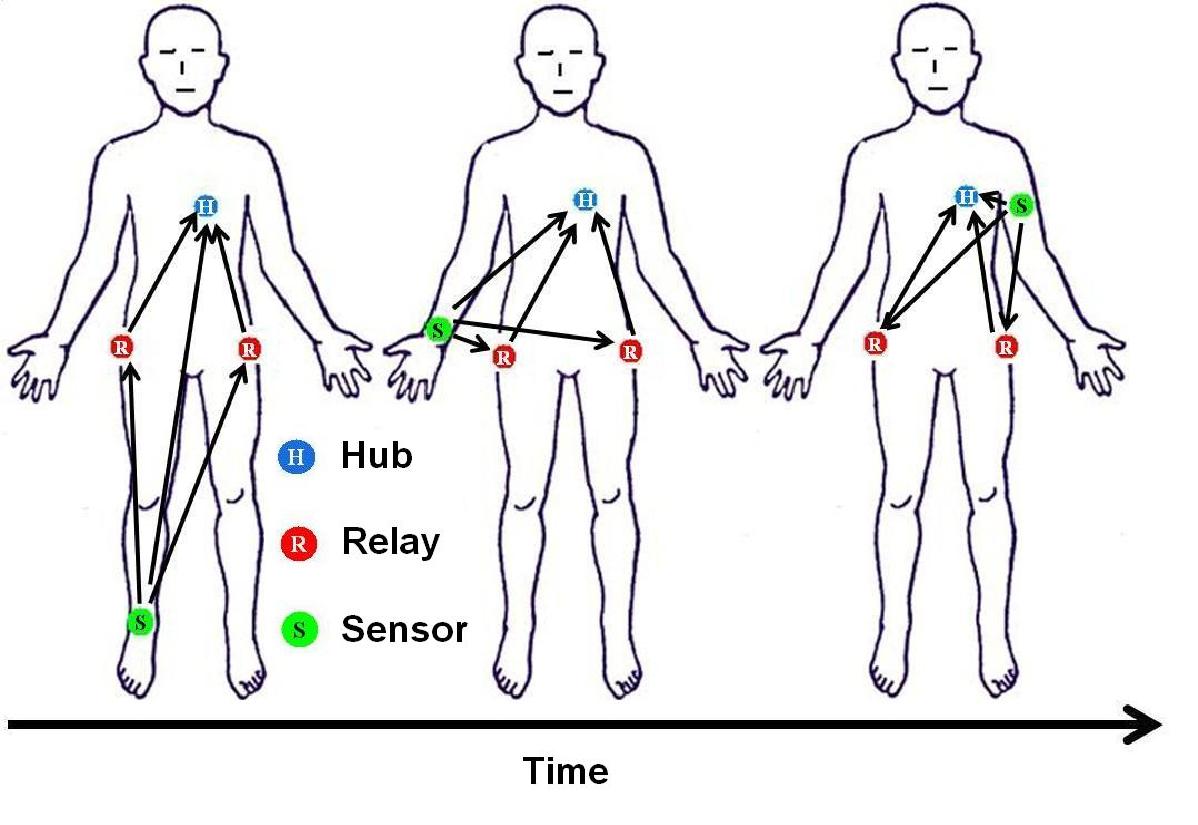}
\label{fig:Fixed Relays Implementation}}
\caption{Two different relay selection implementations}
\label{fig:relay selection}
\end{figure}

Two different relay selection schemes are investigated. The first implementation does not require additional hardware, but uses inactive sensors as relays, as shown in Fig.~\ref{fig:Varying Relays Implementation}. As mentioned before, the WBAN consists of one hub and three active sensors. While one sensor is transmitting, the other two act as possible relays and decode-and-forward received packets to the hub as required. In contrast, for the second implementation in Fig.~\ref{fig:Fixed Relays Implementation}, besides the existing sensors and hub, two fixed relays are added to the overall configuration of the WBAN system. The relays are placed at the left and right hips respectively. When active, they listen to the channel and decode-and-forward packets transmitted by sensors.

\subsubsection{intra- and inter-WBAN channel model}
Extensive on-body and inter-body channel data was measured with small wearable channel sounders operating at 2.36 GHz (close to the 2.4 GHz ISM band) over several hours of normal everyday activity \cite{NICTAdata}. In terms of on-body channel data, experiments were repeated on different individual subjects. In each experiment, subjects wore the same measuring system, which consisted of 3 transceivers and 7 receivers \cite{NICTAdata}\cite{Dong2012}. Throughout the entire experiment, transceivers worked in a round-robin fashion broadcasting every 5 ms at 0 dBm. During one radio transmission, the remaining channel sounders, including the remaining inactive transceivers, recorded the received signal strength indicator (RSSI) upon successful detection of a packet.

\begin{figure}[]
\centering
\includegraphics[width=0.8\columnwidth]{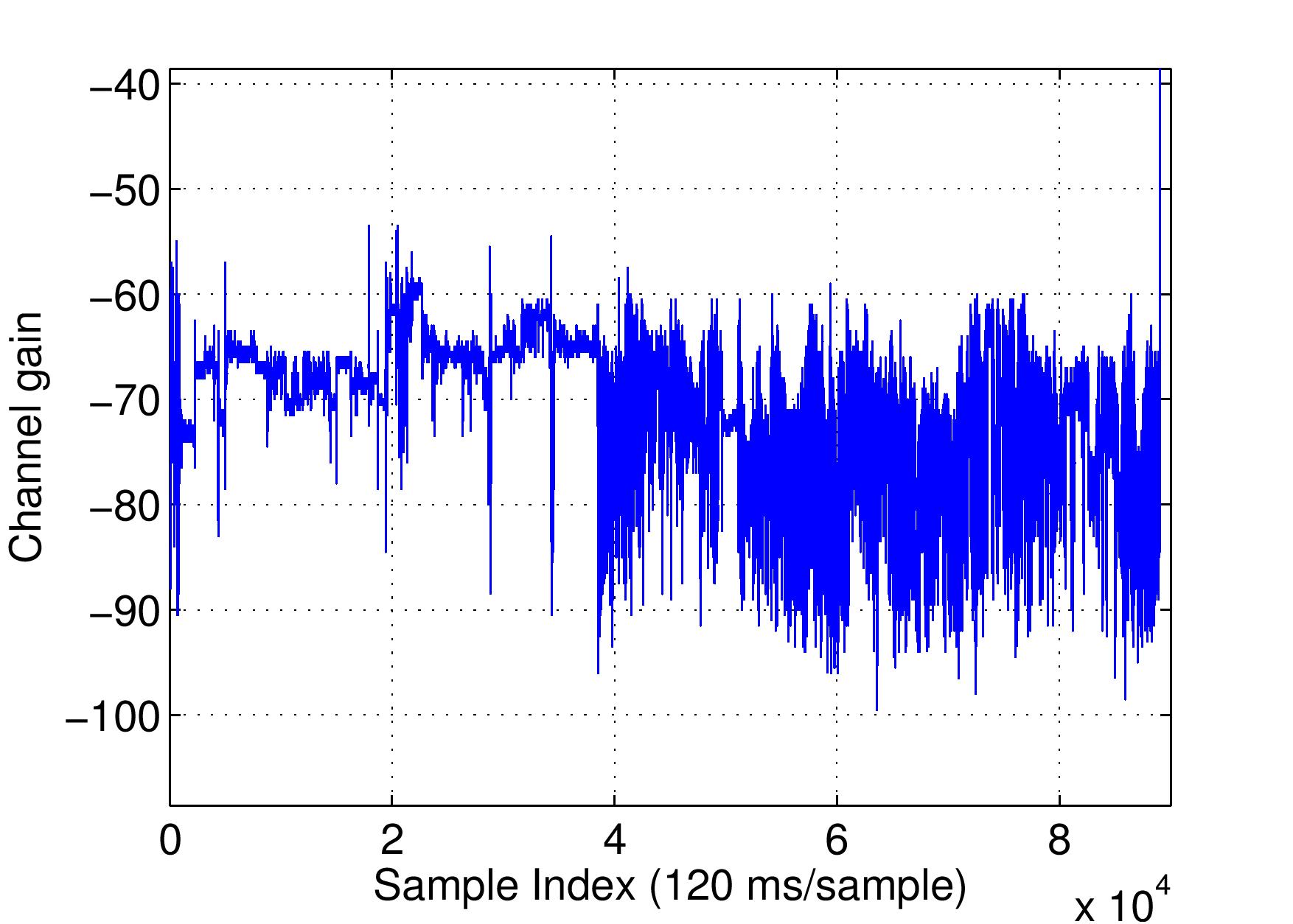}
\caption{A typical on-body channel of Subject 1}
\label{fig: channel_s1}
\end{figure}

While observing different subjects on-body channel profiles, Subject 1's data indicates that these is a sudden change in experiment environment. Fig.~\ref{fig: channel_s1} shows a typical channel gain plot for Subject 1 over a period of approximately 3 hours. As seen in the figure, the sudden change occurs roughly at the 40000th sample (i.e., after approximately 80 minutes). The channel before that moment shows a slow fading characteristic with larger coherence time, and it is more stable. In contrast, after the 40000th sample, the coherence time decreases significantly and the channel starts to vary more rapidly. In the simulation of two-hop cooperative communication scheme with fixed relays, two distinct parts of Subject 1's channel are investigated and the impact of this difference is then compared.

To capture the inter-body channel data, more subjects were involved in one experiment \cite{NICTAdata,Dong:ICC:2013}. For the data set used in our simulation, there were 8 people walking together to a cafe, sitting there for a while and then walking back to office. Each subject wore a measuring system, which consisted of only 1 transceiver at left hip and 2 receivers at right upper arm and left wrist respectively. Similar to the on-body channel experiment, each transceiver broadcasted every 5 ms in round-robin sequence at 0 dBm. However, since the total number of transceivers was different, on-body and inter-body channel data sets have unmatched sampling rates, 15 ms per sample and 40 ms per sample respectively.

\begin{figure}[]
\centering
\includegraphics[width=0.8\columnwidth]{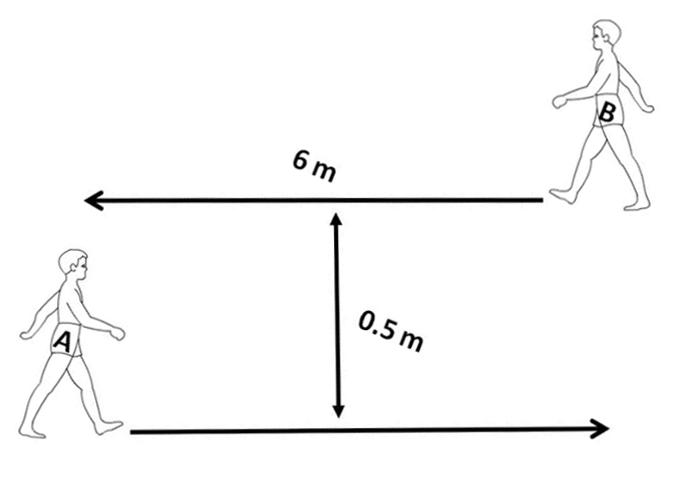}
\caption{Simulation of two subjects' motion for the scenario with variation in relays used}
\label{fig: walking}
\end{figure}

For the analysis with \emph{variation in relays used}, on-body channel data and a simulated inter-body channel model are used, \emph{and hence interference used in this analysis is based on simulation}. It simulates a realistic scenario which may happen in an open indoor or outdoor environment. As shown in Fig.~\ref{fig: walking}, two subjects walk approach each other, passing by each other and then move apart in a space with a dimension of $6\times 0.5~\mathrm{m^2}$. \cite{Dong2012} provides a detailed description of the motion model for this simulation. It also explains how the inter-body channel model is simulated. Briefly, the inter-body channel model incorporates both large scale and small scale fading. The large scale fading model is simulated by combining the effect of free space path loss and shadowing. For WBAN applications, a mean path loss with an exponent of 2 is used since it corresponds to a relatively `open' environment with few objects nearby causing reflection and diffraction. In terms of shadowing, there will not always exist a line-of-sight path between two WBANs. It will be obstructed by the human body and influenced by body movement in most cases. According to the on-body channel data, it can be observed that the average attenuation caused by shadowing is approximately 40~dB. Thus, the large scale fading model is simulated by adding a $-40$ dB offset to the mean path loss. Different levels of shadowing are investigated; no shadowing, partial shadowing and full shadowing, among which full shadowing is the most common. Around shadowed mean path loss, small scale fading is added. This is achieved by using Jakes model with a Doppler frequency of 2Hz and Rayleigh fading with fading coefficients that are $\mathcal{CN}(0,1)$ distributed around the shadowed path loss. \emph{Please note that Rayleigh small-scale fading is a poor model for the on-body channel over the WBAN-of-interest, e.g., \cite{SmithIS08,Smith:AT:2011}, but this model is applied here specifically for an open environment body-to-body.}

\begin{figure}[]
\centering
\includegraphics[width=0.8\columnwidth]{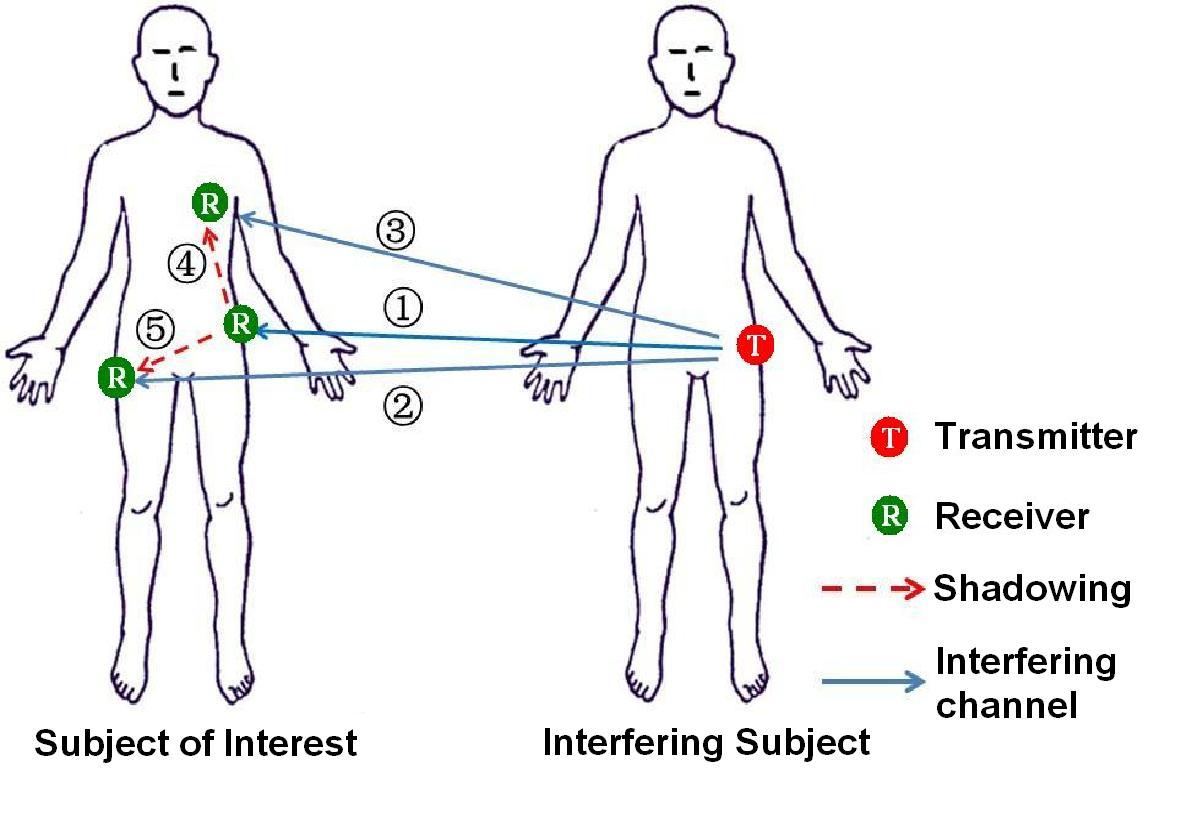}
\caption{Overlaying of inter-body and on-body channels}
\label{fig: Overlaying}
\end{figure}

In the case of analysis with \emph{a fixed relay implementation}, both on-body and inter-body channel measurements are employed, \emph{and hence interference used in this analysis is based on measurement}. Recall that on-body and inter-body sampling rates are different, 15 ms and 40 ms respectively. Therefore, it is essential to synchronize them to the same sampling rate. As opposed to the experiment described in \cite{David:ElecLetter:2009}, in which the experiment involves continuous strenuous activity, the inter-body channel data used in this paper was measured differently. The orientation and distance between subjects were more stable, and there were less scatterers in the surrounding environment. As defined in \cite{Paulraj:2008}, channel coherence time is typically the time lag until the autocorrelation coefficient reduces to 0.7. Based on that, calculations show that on-body and inter-body channel data used in this simulation have an average coherence time of 2087 ms and 892 ms respectively. Therefore, it is possible to down-sample both types of channel data to 120 ms per sample without losing accuracy. In addition, since the transmission time of a packet is shorter than the 120 ms sampling rate, a block fading model is used whereby the Tx-Rx channel gain is constant over each packet. Besides temporal synchronization, spatial overlaying is also required. Since opportunistic relaying used in this simulation is based on the SINR value at the relays and the hub, it needs channel data for all three interfering links. Data for channels 1, 2 and 3 is shown in Fig.~\ref{fig: Overlaying}. However, due to the limitation of the inter-body experiment, there are no direct measurements for Channels 2 and 3. Therefore, they are modeled by overlaying the shadowed components of second hop on-body channels, which are Channels 4 and 5 in Fig.~\ref{fig: Overlaying}, to the existing inter-body channel data. A detailed description of this overlaying process can be found in \cite{Dong:ICC:2013}. Here, different combinations of subjects are used in the overall analysis, as shown in Table \ref{table: Simulation combination}.

\begin{table}[]
\centering
\caption{Choice of WBAN-of-interest and Interfering WBAN: each X indicates an independent analysis set.}
\begin{tabular}{|l||*{6}{c|}}\hline

\multicolumn{1}{|c||}{}&
\multicolumn{6}{|c|}{\textbf{Interfering WBAN}}\\\hline

\textbf{Subject-of-interest}
&\makebox[3em]{\#1}&\makebox[3em]{\#2}&\makebox[3em]{\#3}
&\makebox[3em]{\#4}&\makebox[3em]{\#5}&\makebox[3em]{\#6}\\\whline

Subject \#1 & & x & x & x & x & x \\ \hline
Subject \#2 & x & & x & x & x & x \\ \hline
\end{tabular}
\label{table: Simulation combination}
\end{table}

\section{Simulation Results with Varied Relay Positions}
In this section, the performance of two-hop opportunistic relaying, using (\ref{equ:OR}), with varied relay positions is compared with a system using single-link communications. We also investigate the impact of different hub locations and levels of shadowing on the overall WBAN-of-interest performance. The analysis is based on received SINR outage probability, where the received SINR is derived from recorded signal strength, interference is simulated as described in the last section and noise is AWGN with $|\varepsilon_{\mathrm{noise}}| = -95$~dBm, see (\ref{equ:SINR}). In this paper, the outage probability at a given SINR threshold $\nu_{\mathrm{th}}$ is defined as the probability of the SINR value of a random received packet being smaller than $\nu_{\mathrm{th}}$, i.e. $Pr(\nu \textless \nu_{\mathrm{th}})$. Performance is compared with respect to a SINR threshold value at outage probabilities of 1\% and 10\%, where 10\% corresponds to the guideline for 10\% maximum packet error rates in the IEEE 802.15.6 BAN standard \cite{TRD}.

\begin{figure}[]
\centering
\subfigure[Outage Probability for hub at chest]
{\includegraphics[width=0.6\columnwidth]{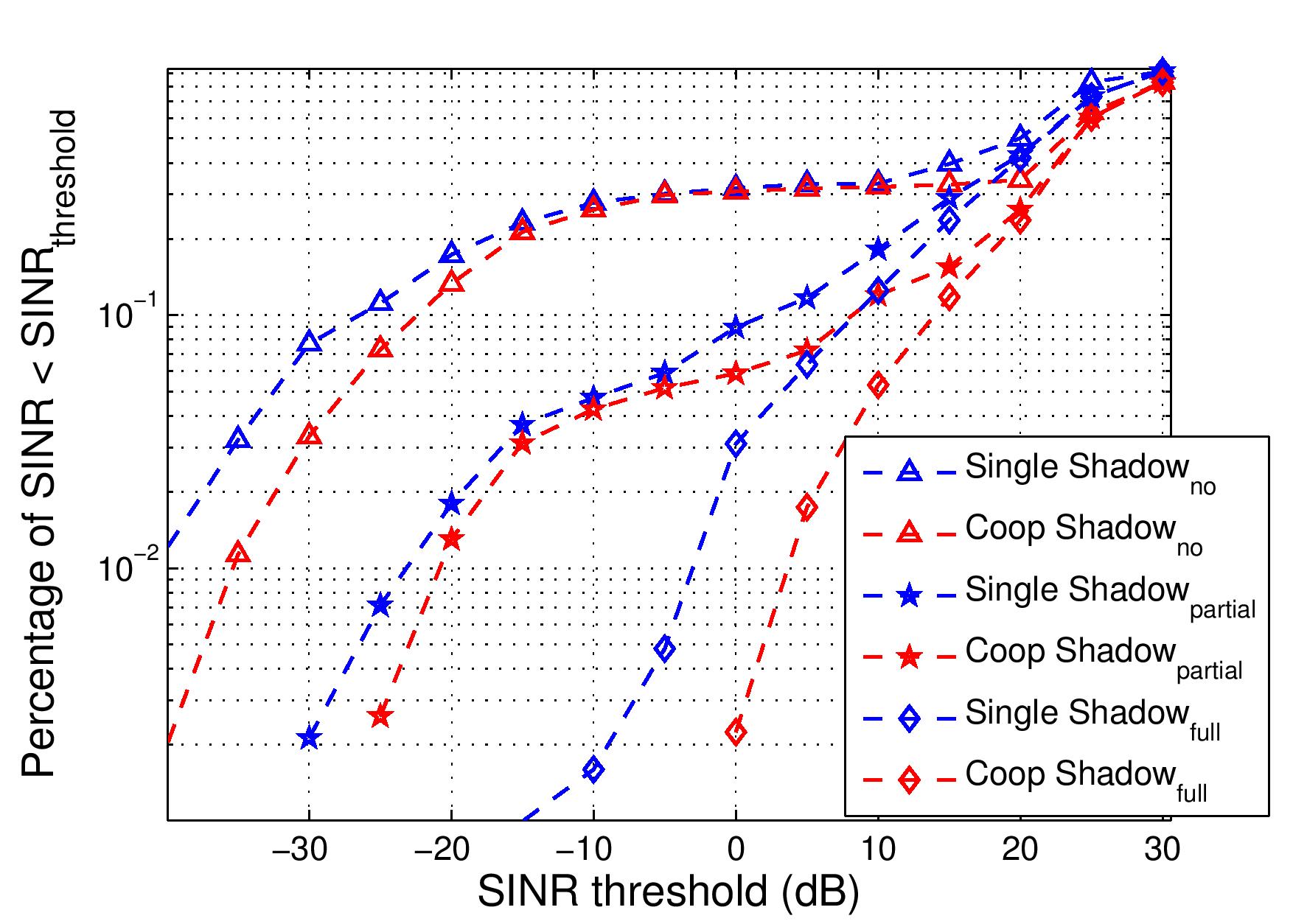}
\label{fig: OP_chest}}
\subfigure[Outage Probability for hub at left hip]
{\includegraphics[width=0.6\columnwidth]{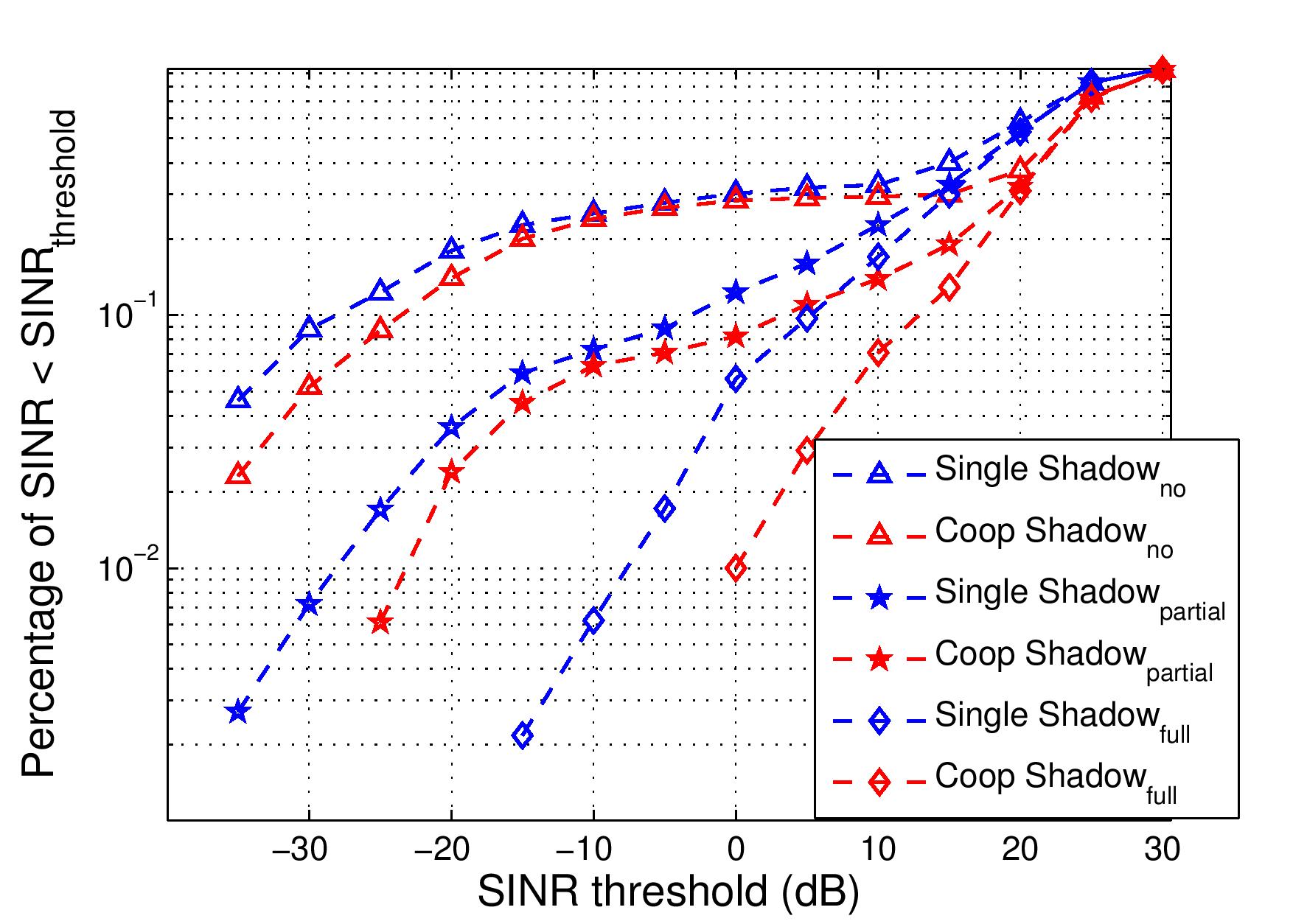}
\label{fig: OP_lefthip}}
\subfigure[Outage Probability for hub at right hip]
{\includegraphics[width=0.6\columnwidth]{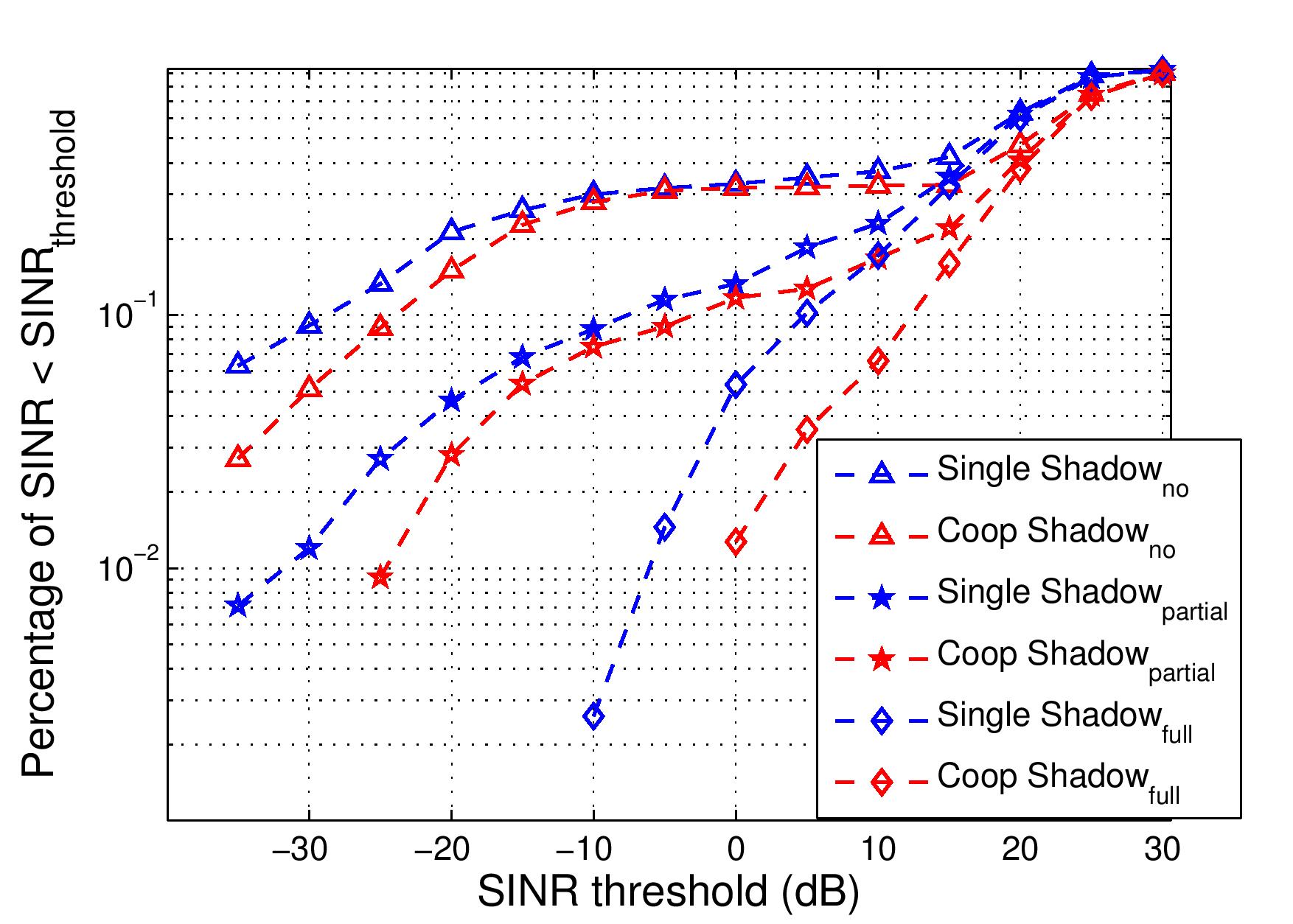}
\label{fig: OP_righthip}}
\caption{SINR Outage Probability for Subjects 1 and 2, varied relay positions.}
\label{fig: outage probability_varyRelay}
\end{figure}

In these simulations, the hub and two sensors are placed at one of the three locations (chest, left and right hips) separately. The last sensor is placed at one of the locations among left and right ankle, left and right wrist, left upper arm, head and back, which correspond to the channel sounder locations in the on-body channel data capturing experiment. In Fig.~\ref{fig: outage probability_varyRelay}, it is obvious that shadowing influences the system performance significantly. In the case of the hub at chest, Fig.~\ref{fig: OP_chest}, partial shadowing raises the SINR threshold value by about 29 dB at an outage probability of 10\% over the case where no shadowing is employed. Full shadowing gives a further 6 dB increment. Similarly, when the hub is placed either at left or right hips, the same observation can be made. Therefore, shadowing is able to mitigate interference from other co-located WBANs, and hence improve the SINR performance for the WBAN-of-interest.

Considering the orientation of interfering hubs or sensors to the WBAN-of-interest, full shadowing is, commonly, a more realistic assumption. Therefore, we compare the performance of two-hop cooperative communications assuming full shadowing. In Fig.~\ref{fig: outage probability_varyRelay}, a significant and consistent improvement in outage probability performance is demonstrated using cooperative communications. Figs. \ref{fig: OP_chest}, \ref{fig: OP_lefthip} and \ref{fig: OP_righthip} show that the proposed cooperative scheme can provide approximately 7~dB improvement at an outage probability of 10\% regardless of the hub location. However, comparing different hub placements, the case where the hub is at the chest shows a better overall performance in terms of outage probability compared with the cases of the hub placed at either the left or right hip.

\section{Theoretical Analysis for Fixed-Relay Scheme}
In this section, analysis is once again based on two-hop relay-assisted communications. For each analysis set marked in Table \ref{table: Simulation combination}, results are collected for each packet. For each packet, received signal and interference power are both recorded, and AWGN with power $|\varepsilon_{\mathrm{noise}}| = -95$~dBm is added. An opportunistic relaying decision using (\ref{equ:OR}) is made according to all calculated SINR values for each packet (\ref{equ:SINR}). Here, the first and second order statistics according to fitted distributions of the data, including SINR distribution, level crossing rate and average outage duration, are obtained.

\subsection{Correlation between Signal and Interference}
Firstly, the cross correlation between signal and interference received at the same destinations is investigated, for all destination links. Received signal and interference powers, $s$ and $int$ respectively, are normalised, and their cross correlation, $\forall i$ packets, is calculated as

 \begin{equation}
\varrho_{s,int}=\frac{E\{\left(s_{i}-E(s)\right).\left(int_i-E(int)\right)\}}{\sqrt{\mathrm{var}(s)}\sqrt{\mathrm{var}(int)}}~.
\label{corr}
\end{equation}

According to the analysis of the fixed-relay data used here the cross correlation, $\varrho_{s,int}$ from (\ref{corr}), between signal and interference for all destination links is averaged and found to be 0.27. As described for autocorrelation in the Section II, if the cross-correlation coefficients were consistently above 0.7, then signal and interference would be significantly correlated.Thus, an average cross-correlation of 0.27 indicates signal and interference are, generally, independent.

Moreover, based on the signal and interference powers recorded, we calculate the best envelope distribution for received signal and interference powers separately, and the best fit is found among normal, lognormal, gamma, Weibull, Nakagami-m and Rayleigh distributions, according to the minimum negative log-likelihood of maximum-likelihood (ML) estimation of distribution parameters. Based on these distribution estimations, we form separate probability density functions, $f_S(s)$ and $f_I(int)$, for signal and interference respectively. We also calculate the empirical joint PDF of both signal and interference for each packet, $f_{S,I}(s,int)$. We confirm the independence of signal and interference as we find that,

\begin{equation}
f_{S,I}(s,int)\simeq f_S(s).f_I(int)~.
\label{SIR distribution}
\end{equation}

\subsection{SINR Distribution}
\begin{figure}[]
\centering
\includegraphics[width=0.6\columnwidth]{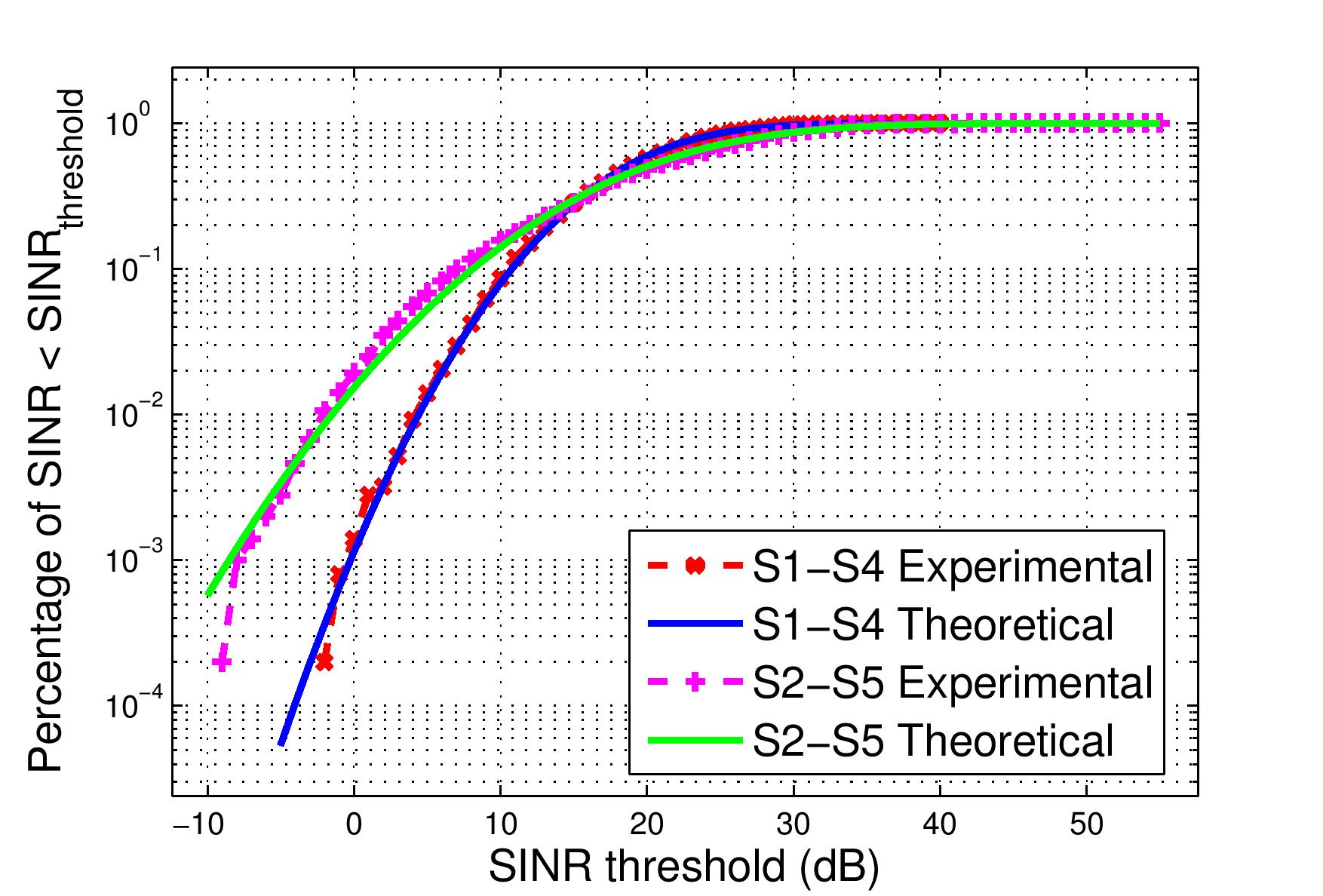}
\caption{Experimental and derived outage probability comparison for WBANs-of-interest on Subjects 1 and 2}
\label{fig:Theoretical OP}
\end{figure}

The distribution of SINR for each simulation using two-hop communications with fixed relays is found according to the ML-Estimation, the same as used for signal and interference powers, described in Section IV.A.  A lognormal best fit for both Subjects 1 and 2 is found when the channel is less stable, while a Nakagami-m distribution is found to best describe the channel with a larger coherence time.  The parameters of best-fitted distribution for each analysis set are shown in Table \ref{table: Distribution}. To see how well the distribution approximation fits the empirical data, the cumulative distribution functions for each results set are derived based on the parameters found using ML-Estimation and then compared with the corresponding outage probability $Pr(\nu < \nu_{th})$. In Fig.~\ref{fig:Theoretical OP}, two examples, Subject 1 with  interfering Subject 4 and Subject 2 with interfering Subject 5, are shown. It can be observed that  there is a very good distribution fit to experimental results.

\begin{table*}[t!]
\centering
\caption{Distribution of the SINR values, Fixed Relay Scheme, across analysis sets.}
\begin{tabular}
{|p{2cm}|c|c|c|c|}\hline
  \multicolumn{1}{|c|}{\textbf{Subject-of-interest}}&
  \multicolumn{1}{|c}{\textbf{Interfering Subject}}&
  \multicolumn{1}{|c|}{\textbf{Distribution}}&
  \multicolumn{1}{|c|}{\textbf{Parameters}}&
  \multicolumn{1}{|c|}{\textbf{Comments}}
  \\\whline

\multirow{5}{*}{\parbox{2in}{Subject \#1}} &
  Subject \#2 & normal & $\mu_n=16.3322, \sigma_n=5.1008$ &
  \multirow{5}{*}{\parbox{1.5in}{On-body channel gain data of Subject 1, from first sample to $\sim$40000~th sample (i.e., first 80 minutes), is very stable, as shown in Fig.~\ref{fig: channel_s1}}}
  \\ \cline{2-4}
& Subject \#3 & Nakagami-m & $m=1.3618, w=254.5702$ &   \\ \cline{2-4}
& Subject \#4 & Nakagami-m & $m=1.6476, w=268.4434$ &   \\ \cline{2-4}
& Subject \#5 & Nakagami-m & $m=1.0556, w=212.0279$ &   \\ \cline{2-4}
& Subject \#6 & Weibull & $k=2.2253, \lambda=15.0594$ &   \\ \whline


\multirow{5}{*}{\parbox{2in}{Subject \#1}} &
  Subject \#2 & lognormal & $\mu=2.1292, \sigma=0.6879$ &
  \multirow{5}{*}{\parbox{1.5in}{Channel gain data starting from $\sim$40000th sample is far less stable than start of dataset, as shown in Fig.~\ref{fig: channel_s1}}}
  \\ \cline{2-4}
& Subject \#3 & gamma & $a=2.5504, b=2.7798$ &   \\ \cline{2-4}
& Subject \#4 & lognormal & $\mu=2.1374, \sigma=0.7004$ &   \\ \cline{2-4}
& Subject \#5 & lognormal & $\mu=1.5759, \sigma=0.7354$ &   \\ \cline{2-4}
& Subject \#6 & lognormal & $\mu=1.5933, \sigma=0.8602$ &   \\ \whline


\multirow{5}{*}{\parbox{2in}{Subject \#2}} &
  Subject \#1 & lognormal & $\mu=2.4652, \sigma=0.9402$ &
  \multirow{5}{*}{\parbox{1.5in}{}}
  \\ \cline{2-4}
& Subject \#3 & lognormal & $\mu=2.3622, \sigma=1.0179$ &   \\ \cline{2-4}
& Subject \#4 & gamma & $a=1.2619, b=12.9156$ &   \\ \cline{2-4}
& Subject \#5 & lognormal & $\mu=2.2883, \sigma=1.0579$ &   \\ \cline{2-4}
& Subject \#6 & lognormal & $\mu=2.5453, \sigma=0.9327$ &   \\ \whline
\end{tabular}
\label{table: Distribution}
$\quad$\\
\footnotesize{$\{\mu_n,\sigma_n\}$, mean and standard deviation of normal distribution; $\{m,w\}$, shape and spread parameters of Nakagami-m distribution;$\{k,\lambda\}$, shape and scale parameters of Weibull distribution; $\{\mu,\sigma\}$, log-mean and log-standard deviation of lognormal distribution; $\{a,b\}$, shape and scale parameters of gamma distribution.}
\end{table*}

\subsection{Level Crossing Rate}

In this paper, level crossing rate (LCR) is defined as the average rate at which a received packet's SINR value going below a particular threshold. In the simulation, the LCR value at threshold $\nu_{th}$ is calculated as

\begin{equation}
  LCR_{\nu_{\mathrm{th}}} = \frac{n}{\sum\limits_{i = 1}^{n-1}{t_{i,i+1}}}~,
\label{equ: LCR}
\end{equation}
where $n$ is the total number of crossings at threshold $\nu_{\mathrm{th}}$, and $t_{\mathrm{i,i+1}}$ indicates the time between two consecutive crossings.

\begin{figure}[]
\centering
\includegraphics[width=0.6\columnwidth]{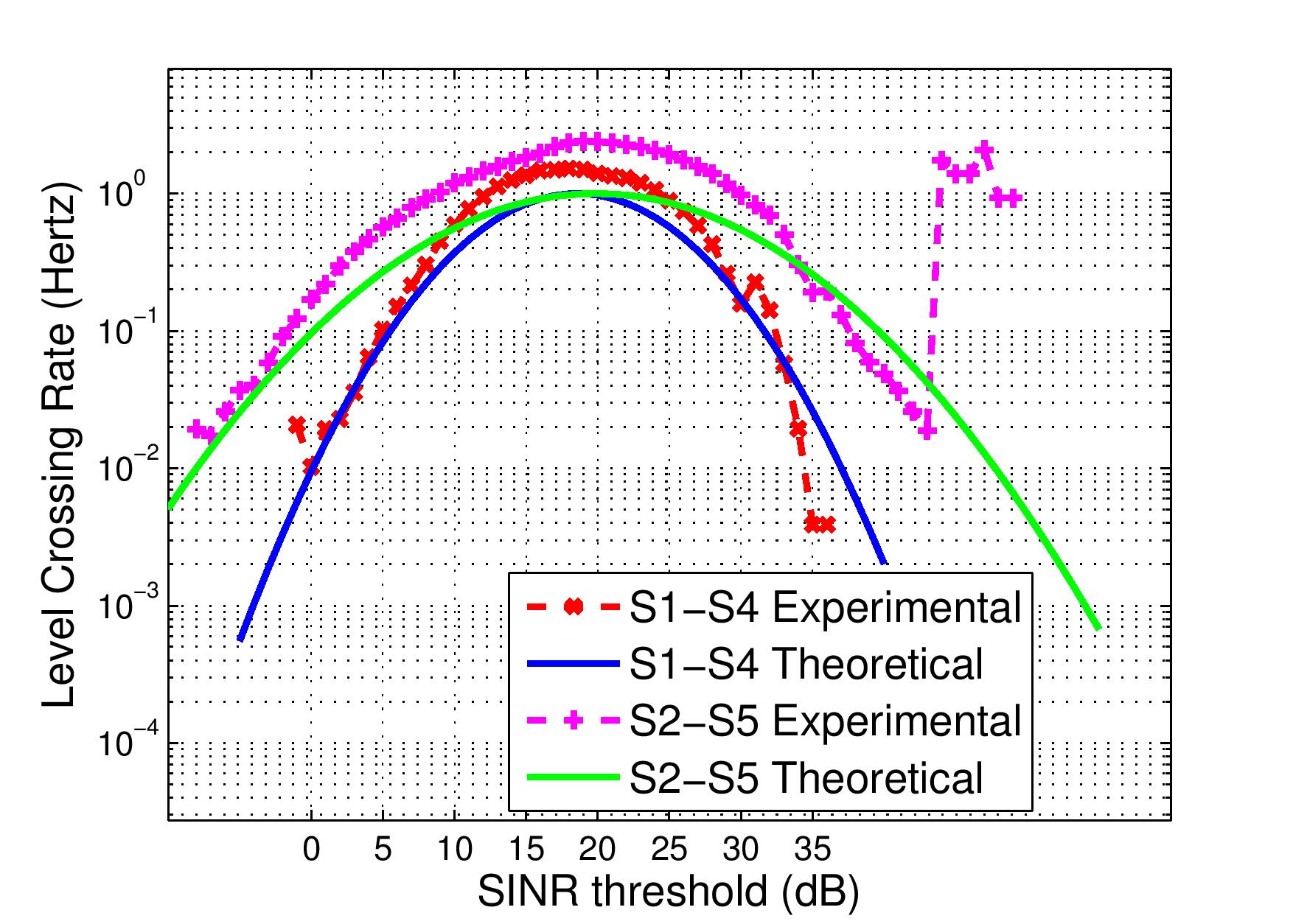}
\caption{Experimental and derived level crossing rate comparison for WBANs-of-interest on Subjects 1 and 2}
\label{fig: Sample Theoretical LCR fit for Subject 1 an 2}
\end{figure}

Here, we also derive the theoretical level crossing rate for every simulation based on the distribution parameters shown in Table \ref{table: Distribution}, and compare with the corresponding experimental LCR. Analysis is performed on simulation for the less stable on-body channel, i.e. Rows 6 -- 15 in Table \ref{table: Distribution}. The theoretical level crossing rate for lognormal and gamma distributions are,

\begin{align}
  LCR_{ln}(\nu_{th}) &= f_D\exp\left(\frac{-(\ln(\nu_{th})-\mu)^2}{2\sigma^2}\right)~, \\
  LCR_{gamma}(\nu_{th}) &= \frac{f_D\sqrt{2\pi}\nu_{th}^{a-0.5}}{\Gamma(a)b^{a-0.5}}\exp(-\nu_{th}/b)~,
\end{align}
where $\nu_{th}$ is the linear value of the threshold; $\ln(\cdot)$ is the natural logarithm and $\Gamma(\cdot)$ is the standard gamma function and $f_D$ indicates the Doppler spread which is found to be 1.0~Hz. It is found that the theoretical LCR curves match the experimental results for all simulations with different subjects-of-interest and interfering subjects. Examples are demonstrated as shown in Fig.~\ref{fig: Sample Theoretical LCR fit for Subject 1 an 2}. Therefore, the fitted distribution and their ML-estimated parameters provide very good estimates for the first and second-order statistics of SINR.

\subsection{Average Outage Duration}
Average outage duration (AOD), analogous to average fading duration for channel gain profiles, is defined as the average period of continuous time that SINR values stay below a threshold value $\nu_{\mathrm{th}}$ across one or more packets. Assume in a given duration, there are a total of $n\geq1$ periods when received packets have SINR values lower than $\nu_{\mathrm{th}}$ and the length of each period is denoted as $t_i$, then the corresponding AOD is calculated as

\begin{equation}
AOD_{\nu_{th}} = \frac{\sum\limits_{i = 1}^{n}{t_{i}}}{n}~.
\label{equ:Average fading duration}
\end{equation}

\begin{figure}[]
\centering
\includegraphics[width=0.6\columnwidth]{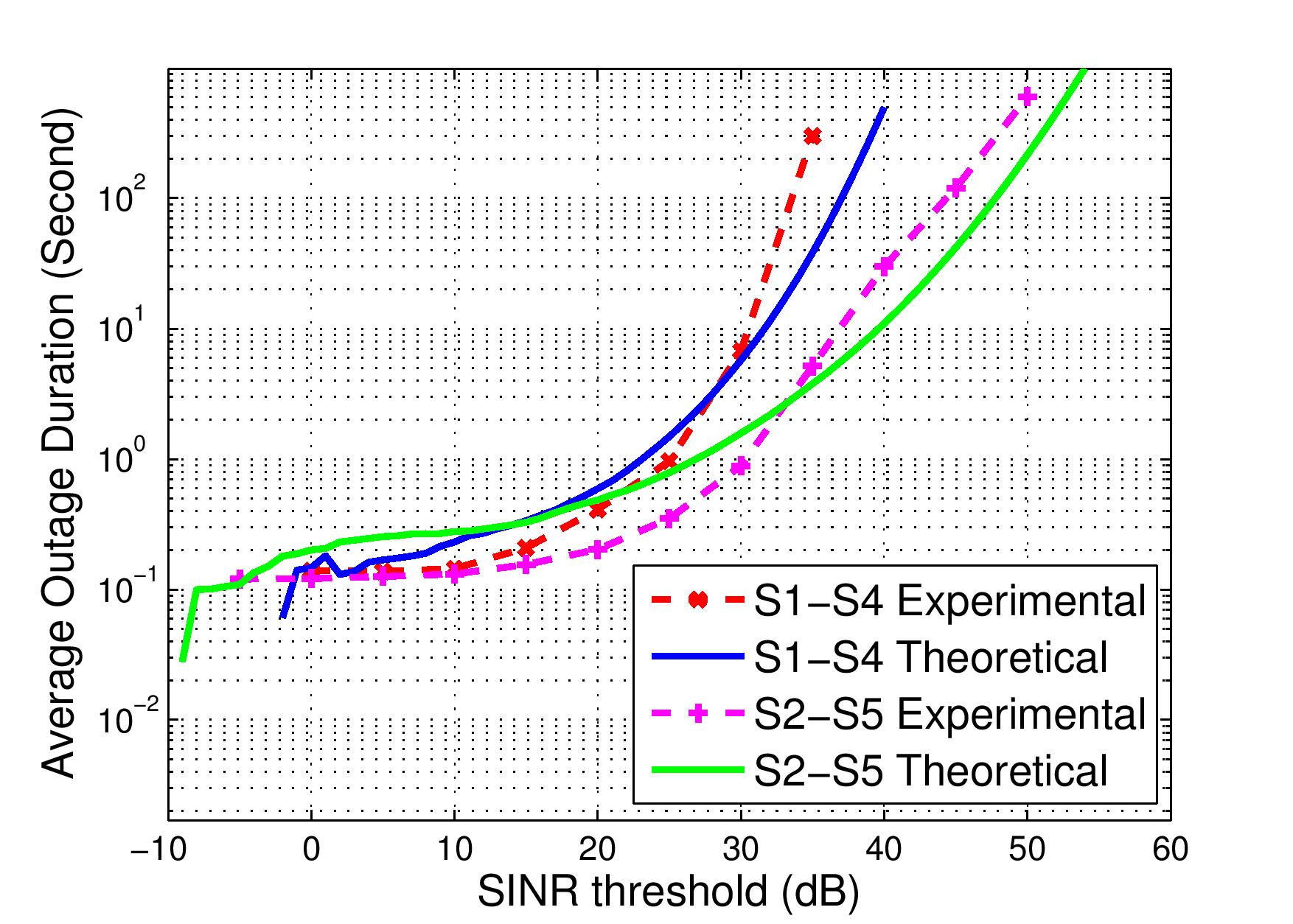}
\caption{Experimental and derived average outage duration comparison for WBANs-of-interest on Subjects 1 and 2}
\label{fig:Theoretical AOD}
\end{figure}

The experimental AOD result is then compared with corresponding theoretical AOD, which is derived as

\begin{equation}
  AOD_{theo}(\nu_{th}) = \frac{F(\nu_{th})}{LCR_{\nu_{th}}}~,
\end{equation}
where $F(\nu_{th})$ is the cumulative distribution function of the SINR values, which is also known as the outage probability at given SINR threshold; $LCR_{\nu_{\mathrm{th}}}$ is the theoretical level crossing rate for the same simulation. From the examples shown in Fig.~\ref{fig:Theoretical AOD}, we can see that theoretical AOD matches experimental results very well.

\section{Simulation Results for the Fixed Relay Scheme}
In this section, the experimental performance of two-hop opportunistic relaying cooperative communications using relays at left and right hips is compared with the system using single-link communications. The analysis is presented, with respect to the SINR values across packets, for the system's first order statistics --- in terms of outage probability; and second order statistics --- in terms of level crossing rate and average outage duration. Simulation results for using Subjects 1 and 2 as subject-of-interest are shown separately. In addition, on-body channels of Subject 1 shows two distinct characteristics after a sudden change of experiment environment, as shown in Fig.~\ref{fig: channel_s1}. Channels are very stable at an earlier stage, i.e., first 80 minutes, and then they become less stable. Therefore, the performance of the proposed scheme in these two circumstances is also presented.

\subsection{Experimental Outage Probability}

\begin{figure}[]
\centering
\subfigure[Outage Probability for Subject 1, less stable channel]
{\includegraphics[width=0.6\columnwidth]{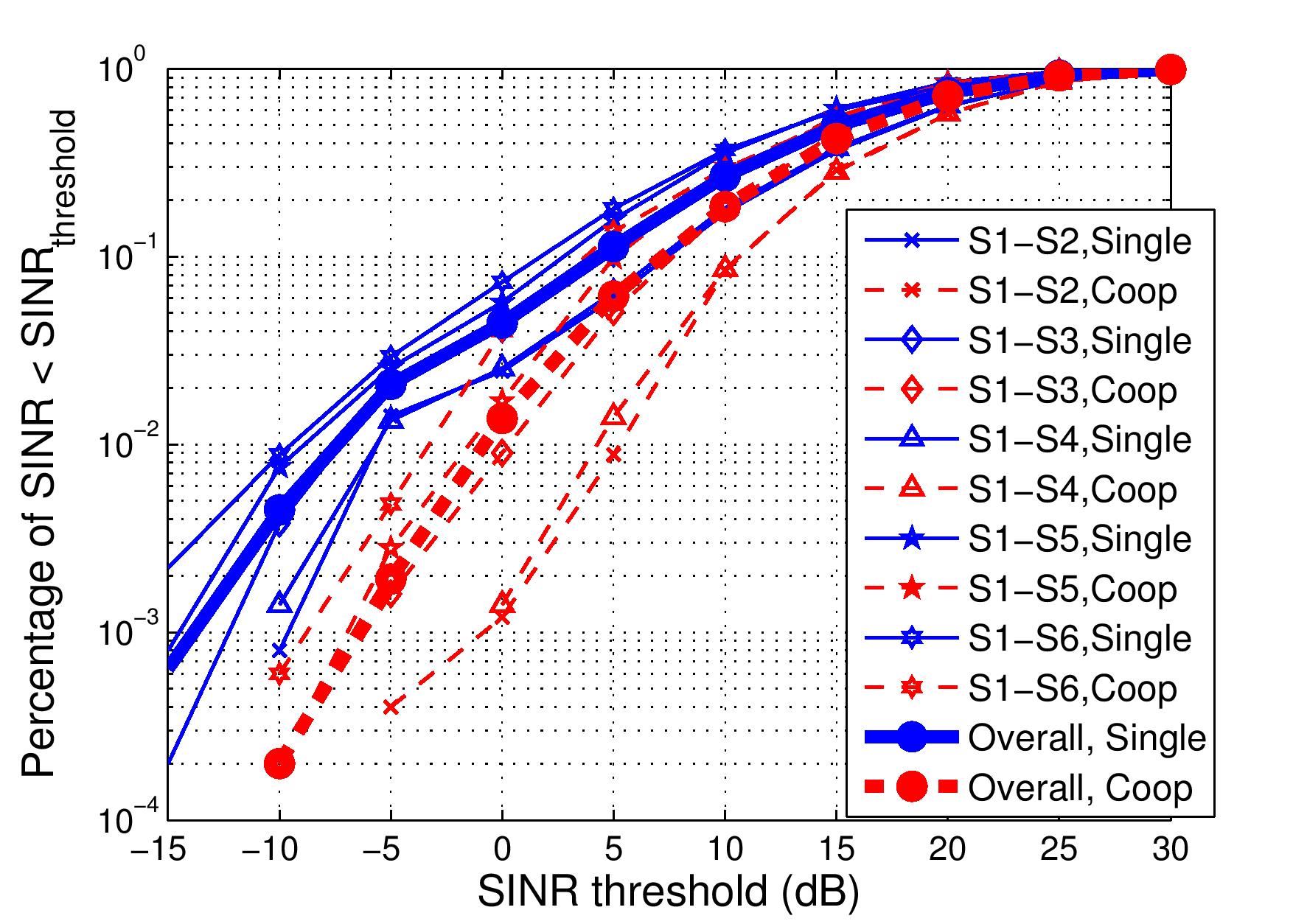}
\label{fig: OP_s1_fast}}
\subfigure[Outage Probability for Subject 2]
{\includegraphics[width=0.6\columnwidth]{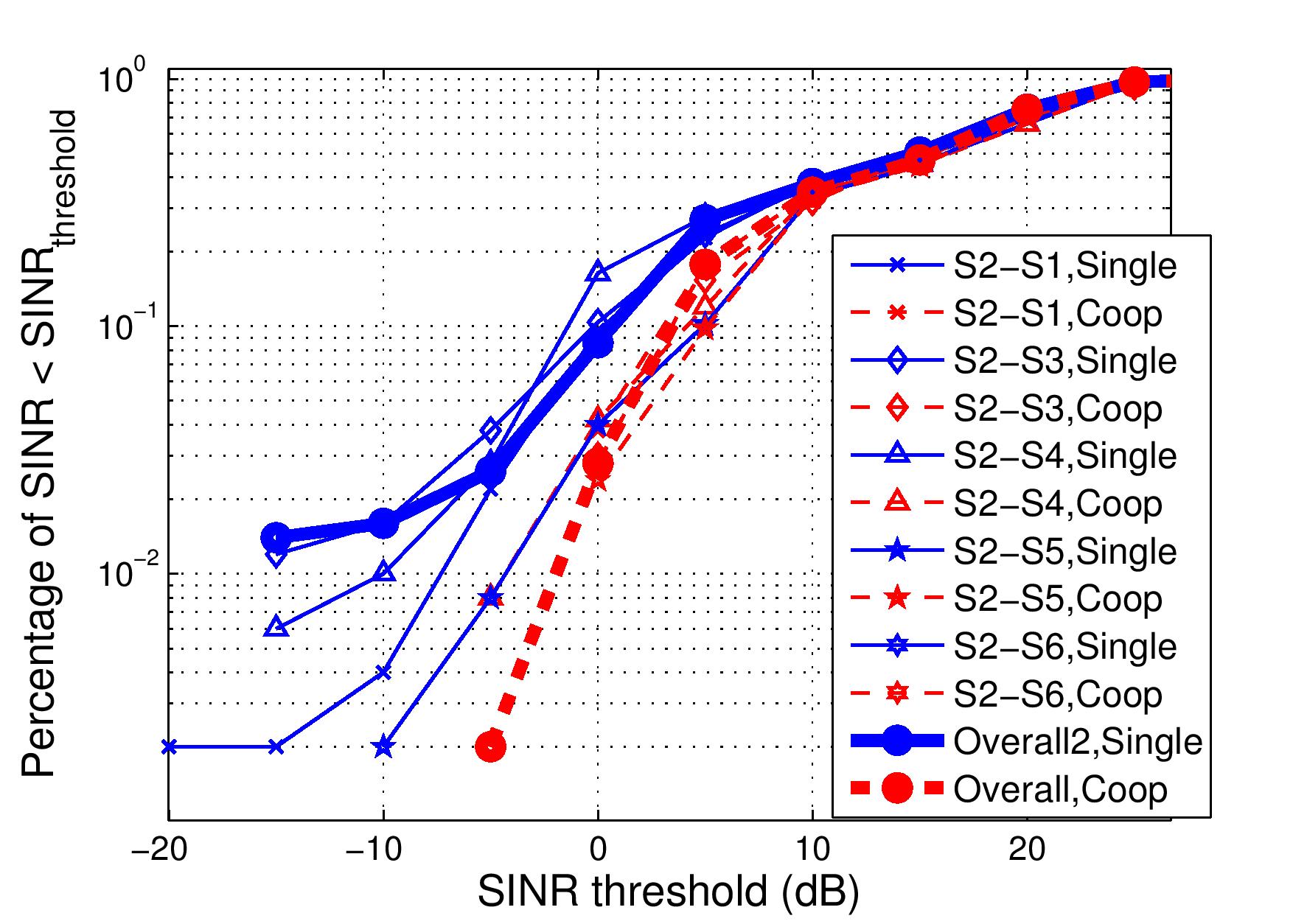}
\label{fig: OP_s2}}
\subfigure[Outage Probability for Subject 1, very stable channel]
{\includegraphics[width=0.6\columnwidth]{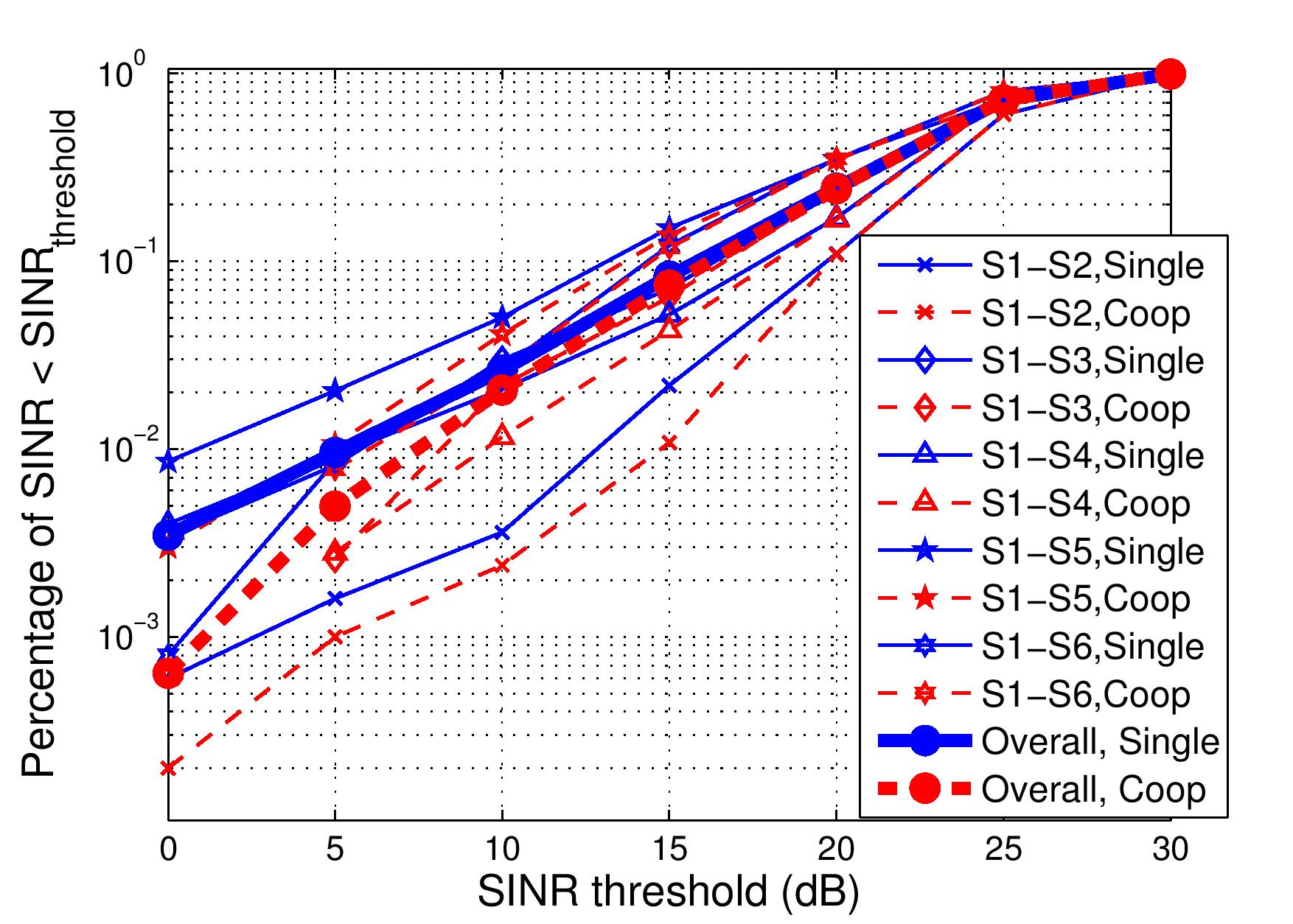}
\label{fig: OP_s1_slow}}
\caption{SINR Outage Probability for Subjects 1 and 2, Fixed Relay Scheme}
\label{fig: outage probability}
\end{figure}

Fig.~\ref{fig: OP_s1_fast} - \ref{fig: OP_s1_slow} shows the resultant outage probability for Subjects 1 and 2. In the figures, the overall outage probability is obtained by averaging across the entire set of outage probability for each subject-of-interest. For Subject 1 when the channel is less stable, it is shown in Fig.~\ref{fig: OP_s1_fast} that cooperative communications provides an average of 3 dB and 7 dB improvement over single link communications at outage probabilities of 10\% and 1\% respectively. Looking at individual analysis sets, the improvement can reach up-to 11 dB increase in threshold value at an outage probability of 1\% when Subject 2 is taken as an interferer. In addition, simulation on Subject 2 shows similar improvements, with 3 dB and more than 10 dB increase at 10\% and 1\% respectively (As shown in Fig.~\ref{fig: OP_s2}).

In contrast, simulations on Subject 1 with a very stable channel leads to a different result. In Fig.~\ref{fig: OP_s1_slow}, it is shown that the cooperative communications doesn't provide significant performance improvement in terms of outage probability. However, while comparing with Fig.~\ref{fig: OP_s1_fast} where there is a less stable channel, the system in a very stable channel has a 10 dB  performance gain, for both two hop cooperative and single-link communication schemes, over the less stable channel.

\subsection{Experimental Level Crossing Rate}
\begin{figure}[]
\centering
\subfigure[Level Crossing Rate for Subject 1, less stable channel]
{\includegraphics[width=0.6\columnwidth]{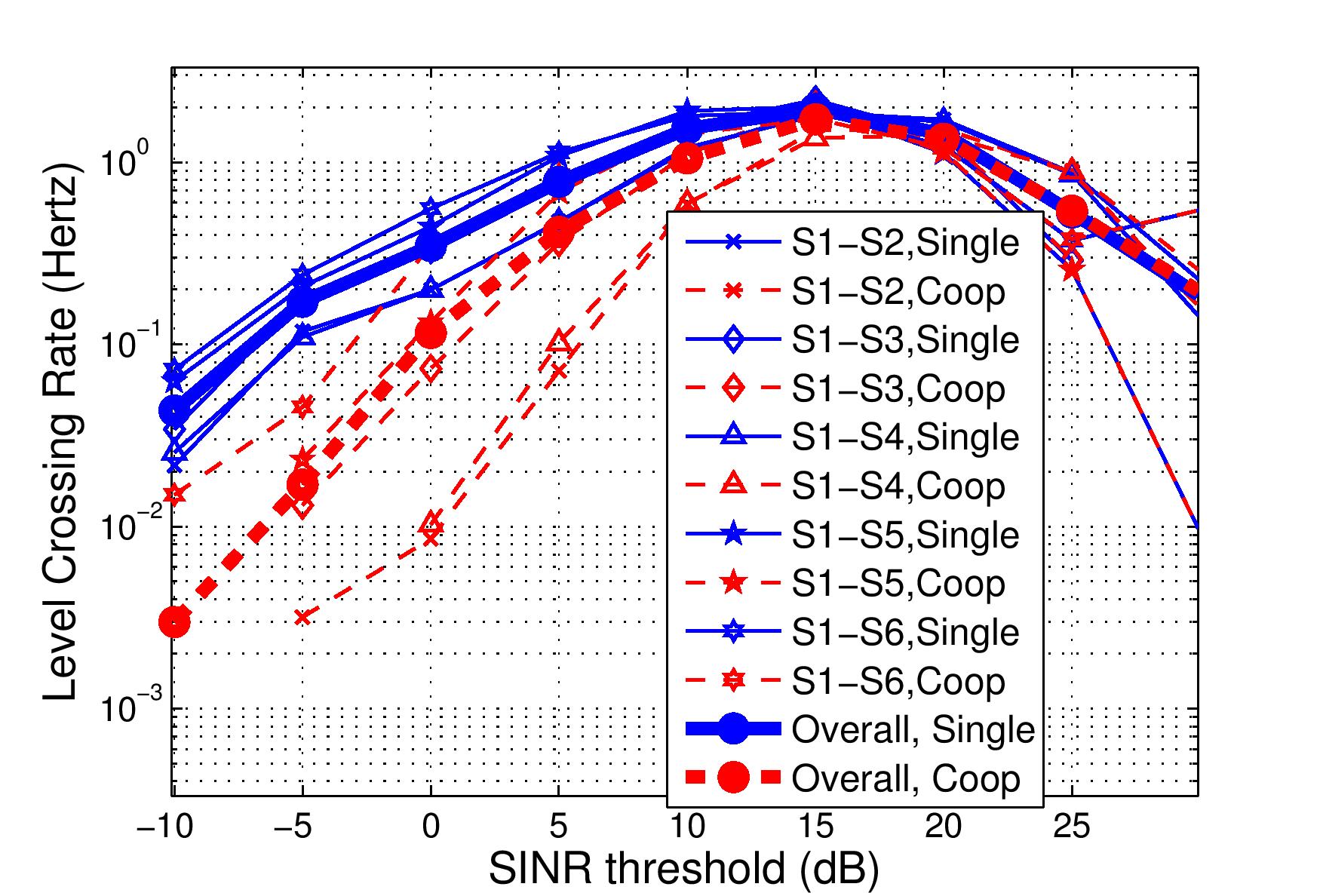}
\label{fig: LCR_s1_fast}}
\subfigure[Level Crossing Rate for Subject 2]
{\includegraphics[width=0.6\columnwidth]{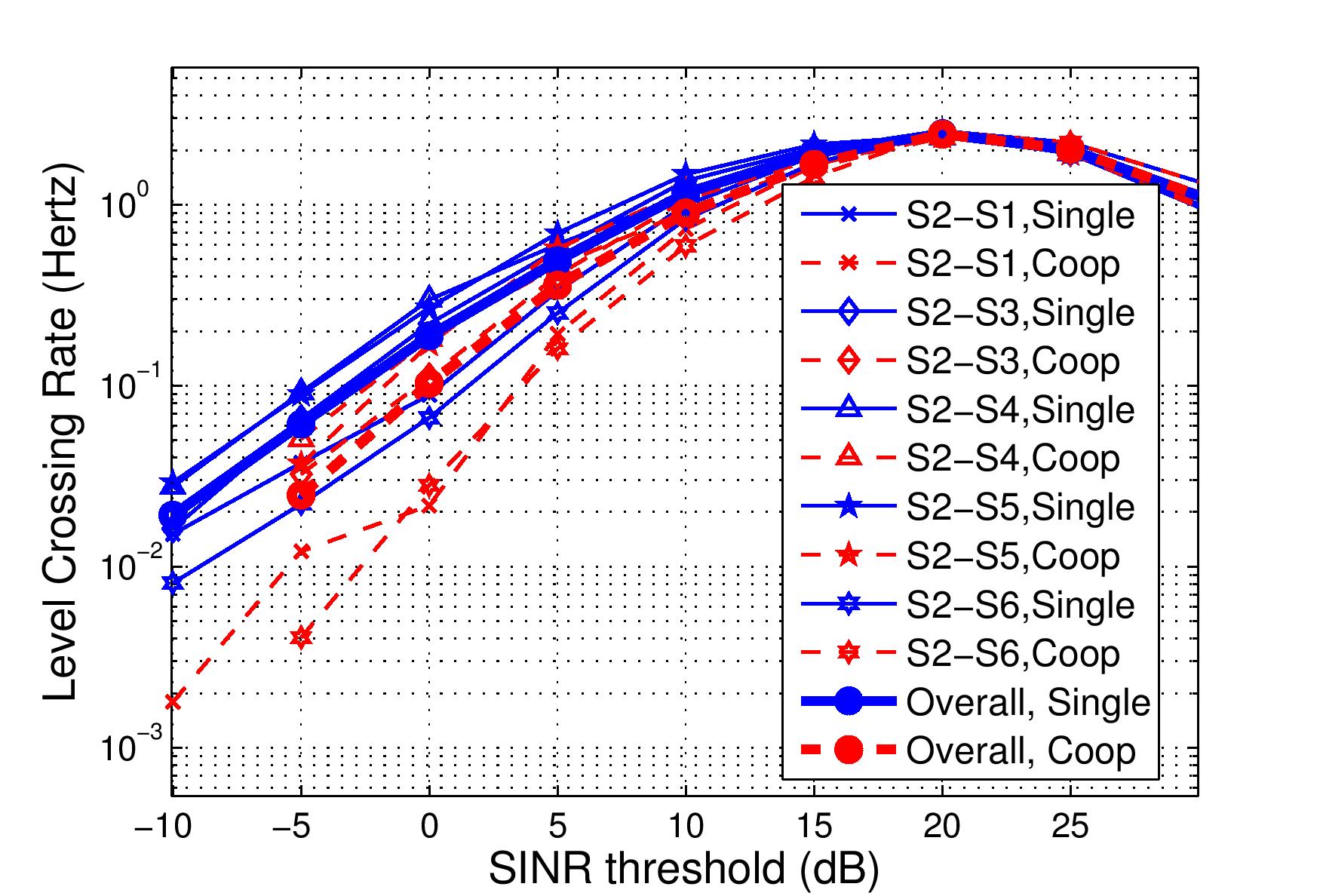}
\label{fig: LCR_s2}}
\subfigure[Level Crossing Rate for Subject 1, very stable channel]
{\includegraphics[width=0.6\columnwidth]{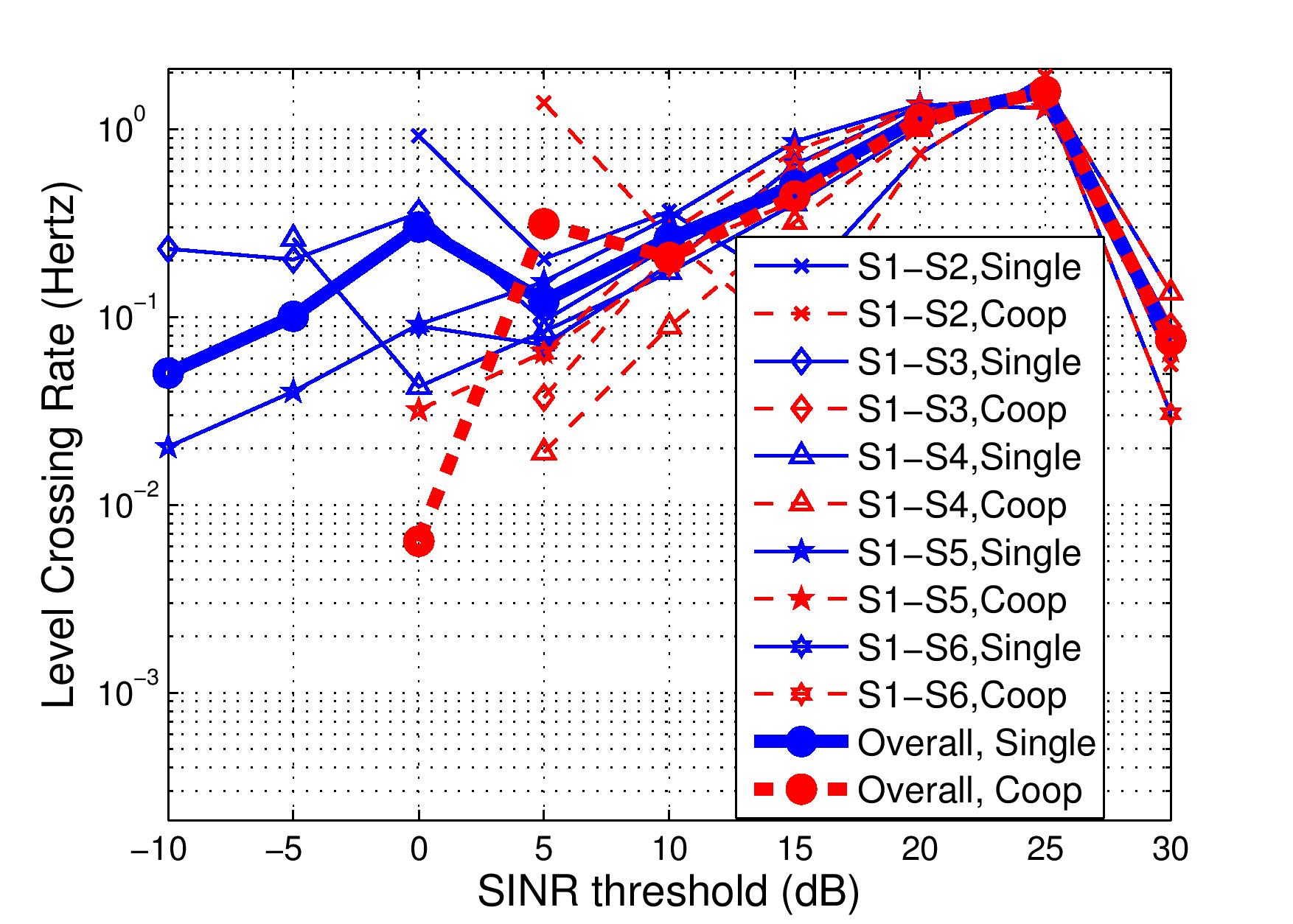}
\label{fig: LCR_s1_slow}}
\caption{Level Crossing Rate of SINR for Subjects 1 and 2,  Fixed Relay Scheme}
\label{fig: level crossing rate}
\end{figure}

As defined in Section IV.C, level crossing rate (LCR) is the average rate of a received packet's SINR value going below a given threshold $\nu_{\mathrm{th}}$. It is calculated as (\ref{equ: LCR}). In Fig.~\ref{fig: LCR_s1_fast} and Fig.~\ref{fig: LCR_s2}, it is shown that cooperative communications reduces LCR at low SINR threshold values significantly. For Subject 1 with less stable channels, the proposed scheme raises the threshold value by an average of 4 dB at an LCR of 1 Hz and an average of 7.5 dB at an LCR of 0.1 Hz. In terms of Subject 2, the improvement, as shown in Fig.~\ref{fig: LCR_s2}, is about 2.5 dB and 3 dB at an LCR of 1 Hz and 0.1 Hz respectively. Furthermore, the same observation as that made for outage probability can be made from Fig.~\ref{fig: LCR_s1_slow} when on-body channels are more stable. In this case of larger channel coherence time, no real performance advantage in terms of level crossing rate is shown for using cooperative communication scheme over single-link communication schemes.

\subsection{Experimental Average Outage Duration}
\begin{figure}[]
\centering
\subfigure[Average Outage Duration for Subject 1]
{\includegraphics[width=0.6\columnwidth]{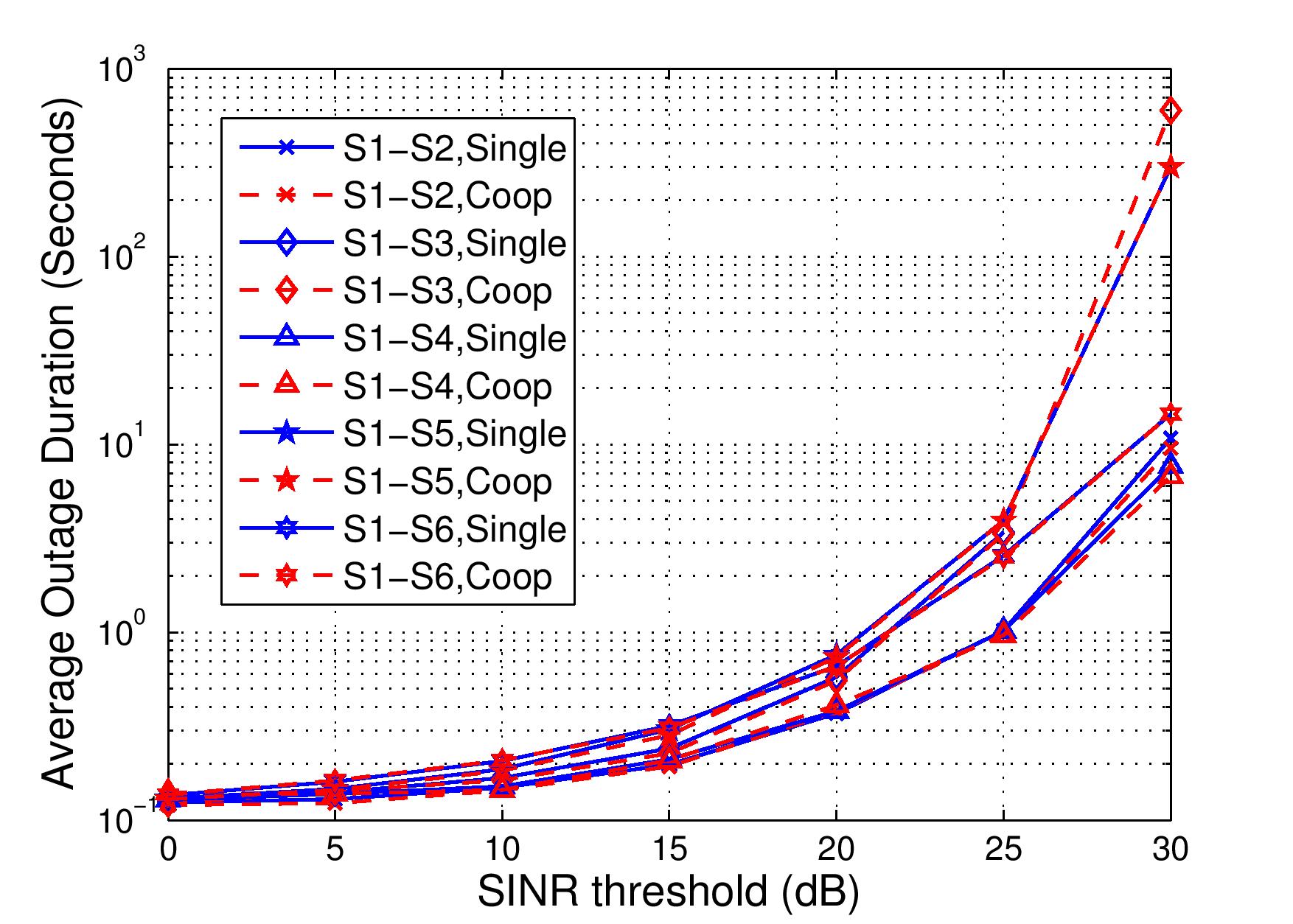}
\label{fig: ADO_s1}}
\subfigure[Average Outage Duration for Subject 2]
{\includegraphics[width=0.6\columnwidth]{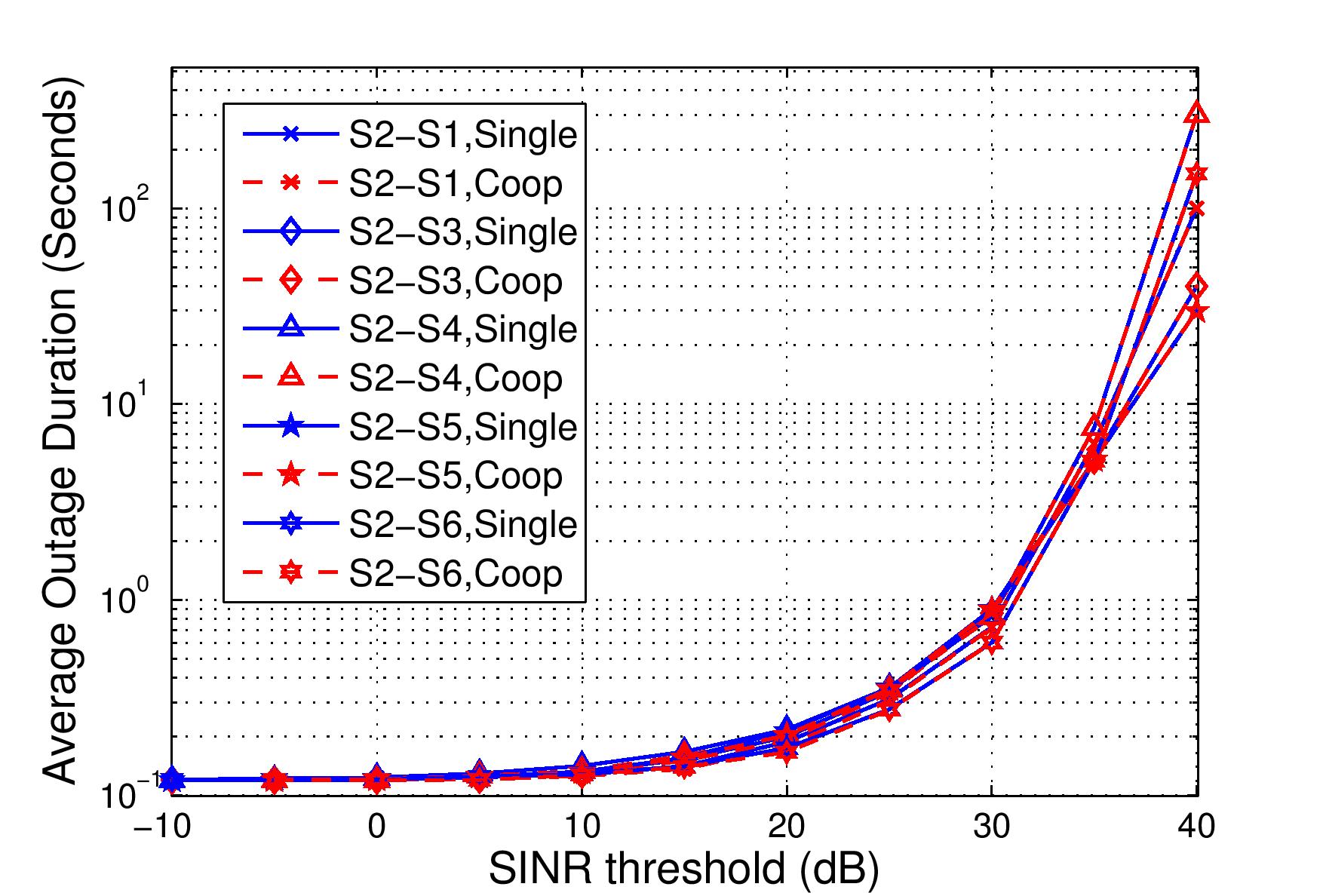}
\label{fig: ADO_s2}}
\caption{Average Outage Duration of SINR for Subjects 1 and 2,  Fixed Relay Scheme}
\label{fig:average outage duration}
\end{figure}

Average outage duration (AOD) of time varying SINR, is calculated as (\ref{equ:Average fading duration}). Fig.~\ref{fig: ADO_s1} and Fig.~\ref{fig: ADO_s2} presents the simulation results of AOD of both corresponding subjects-of-interest and interfering subjects. For both of them, the curves for single-link and two-hop cooperative communication schemes overlap at most of the SINR threshold values. In other words, there is no real performance advantage of using the proposed scheme for WBANs coexistence in terms of average outage duration.

\section{Conclusion}
This paper has studied the performance of two-hop opportunistic relaying cooperative communications with respect to interference mitigation and coexistence enhancement. TDMA was employed as the intra-network access scheme, as well as the access scheme across multiple WBANs. Three-branch opportunistic relaying was used and the best path was selected between an active sensor and the hub. Two different relay implementations were investigated to provide diversity gain at the hub. The first one uses two inactive sensors as relays, while the other implementation adds two additional fixed relays, not providing sensor functions, at the left and right hips. Empirical on-body channel measurements were used as a reliable and realistic intra-WBAN channel model for both implementations. However, there were two implementations of inter-WBAN channel models. For the implementation of varied relay positions, it was simulated by the superposition of free space path loss, shadowing and small scale fading. In terms of simulation using fixed relays, empirical inter-body channel measurements were used in the inter-WBAN channel model of a practical WBAN working environment. The performance of both schemes were compared with traditional single-link communications where no relays are used. The analysis was performed with respect to first and second order statistics of received packets' SINR values. In addition, distributions of SINR values were found using maximum-likelihood (ML) estimation. Theoretical outage probability, level crossing rate and average outage duration were derived based on the distribution parameters, and shown to match empirical results well.

When using the varied relay positions implementation, it has been found that the use of cooperative communications can mitigate interference by providing an average of 7 dB improvement at SINR outage probability of 10\% over the use of single-link communications. This improvement is consistent no matter where the location of the hub is, but better overall performance is obtained when the hub is placed at the chest. In addition, body shadowing can assist WBAN coexistence significantly. In terms of using the implementation with fixed relay positions, similar conclusions on the the effectiveness of cooperative communications can be made. The average improvements in outage probability at 10\% and 1\% are 3 dB and 8 dB respectively. Level crossing rate reduces significantly at low SINR threshold values when using the proposed scheme. However, it does not provide improvement for coexistence in terms of average outage duration of SINR. Based on ML estimation, it is found that a lognormal distribution best describes the probability distribution of the received packets' SINR when the channel coherence time is smaller. In contrast, a Nakagami-m distribution fits SINR better when the channel coherence time is very large, i.e. the channel is highly stable.

\section*{Acknowledgment}

NICTA is funded
by the Australian Government as represented by the Department of
Broadband, Communications and the Digital Economy and the Australian
Research Council through the ICT Centre of Excellence program.

\ifCLASSOPTIONcaptionsoff
  \newpage
\fi

\FloatBarrier




\end{document}